\documentclass{./jfm}

\usepackage{graphicx}
\usepackage{epstopdf, epsfig}
\usepackage[dvipsnames]{xcolor}  
\usepackage{hyperref} 
\usepackage{subcaption} 
\usepackage{lipsum}                     
\usepackage{xargs}                      
\usepackage{color} 
\usepackage{comment} 
\usepackage{amsmath} 
\usepackage{cases}
\usepackage[normalem]{ulem}
\usepackage[english]{babel}
\usepackage{blindtext}
\usepackage[textwidth=0.7in,
  textsize=tiny,
  linecolor=OliveGreen!25,
  backgroundcolor=OliveGreen!25,
  bordercolor=OliveGreen!25]{todonotes} 

\usepackage{pgfplots}
\pgfplotsset{compat=1.10}
\usepgfplotslibrary{fillbetween}
\usetikzlibrary{patterns}

\reversemarginpar

\newcommand{\Pro}{\mathcal{P}}

\newcommand{\chevron}[1]{\left\langle #1 \right\rangle}
\newcommand{\pderiv}[2]{\frac{\partial #1}{\partial #2}}
\newcommand{\bigoh}[1]{\mathcal{O}\left(#1\right)}

\newcommand{\retau}{$Re_{\tau}$}

\newcommand{\blue}[1]{{\color{blue} #1}}

\definecolor{green}{RGB}{0,120,0} 
 
\def\A180{{%
    \setbox0\hbox{------}%
    \rlap{\hbox to \wd0{\hss\footnotesize$\bigcirc$\hss}}\box0
}}

\def\B550{{%
    \setbox0\hbox{------}%
    \rlap{\hbox to \wd0{\hss$\square$\hss}}\box0
}}

\def\C1000{{%
    \setbox0\hbox{------}%
    \rlap{\hbox to \wd0{\hss$\triangledown$\hss}}\box0
}}

\def\D2000{{%
    \setbox0\hbox{------}%
    \rlap{\hbox to \wd0{\hss$\triangle$\hss}}\box0
}}

\def\E5200{{%
    \setbox0\hbox{------}%
    \rlap{\hbox to \wd0{\hss$\bigcirc$\hss}}\box0
}}

\title{Near wall patch representation of wall bounded turbulence}
\author{Sean P. Carney\aff{1}, Bj\"{o}rn Engquist\aff{1,2}, and Robert D. Moser\aff{2,3}\corresp{\email{rmoser@oden.utexas.edu}}}
\shortauthor{S.P. Carney, B. Engquist, and R.D. Moser}
\affiliation{\aff{1} Department of Mathematics, The University of Texas at Austin, TX 78712, USA  
\aff{2} Oden Insititute for Computational Engineering and Sciences, The University of Texas at Austin, TX 78712, USA
\aff{3} Department of Mechanical Engineering, The University of Texas at Austin, TX 78712, USA
}

\begin{document}

\maketitle

\begin{abstract}
Recent experimental and computational studies indicate that near wall turbulent
flows can be characterized by universal small scale autonomous dynamics 
that are modulated by large scale structures. 
We formulate numerical simulations of near wall turbulence in a small 
domain localized to the boundary, 
whose size scales in viscous units. 
To mimic the environment in which the near wall turbulence evolves,
 the formulation accounts for the flux of mean momentum through 
the upper boundary of the domain.
Comparisons of the model's two dimensional energy spectra and low order 
single-point statistics with the corresponding quantities computed 
from direct numerical simulations
indicate that it successfully captures the dynamics of the small scale near wall 
turbulence. 

\end{abstract}


\section{Introduction}
High Reynolds number wall bounded turbulent shear flows are characterized by a separation of scales
between the flow in the near-wall region, in which mean viscous stresses play an important
role, and the flow farther away from the wall, where mean viscous effects
are largely negligible. This separation of scales is quantified by the friction
Reynolds number $Re_{\tau} = \delta/\delta_{\nu}$, where $\delta$ is the 
characteristic length scale of the shear layer, such as a channel
half-width, a pipe radius, or a boundary layer thickness, and 
$\delta_{\nu} = \nu/u_{\tau}$ is the viscous length scale, where
$\nu$ is the kinematic viscosity of the fluid,
$u_{\tau} = \sqrt{\tau_w/\rho}$, $\tau_w$ is the mean 
wall shear stress, and $\rho$ is the fluid density. Both the direct numerical 
simulation (DNS) and large eddy simulation (LES) 
 of such wall bounded turbulent flows are expensive, as the 
spatial degrees of freedom required to resolve 
the near-wall layer scale as $\mathcal{O}(Re_{\tau}^{2.5})$ and
$\mathcal{O}(Re_{\tau}^{2})$ for DNS and LES, respectively
\citep{Mizuno:2013dc}. For a large class of flows of technological importance, 
this cost is prohibitive, even on modern high-performance
computing systems. 

\begin{figure}
  \begin{center}
    \includegraphics[width=0.9\textwidth]{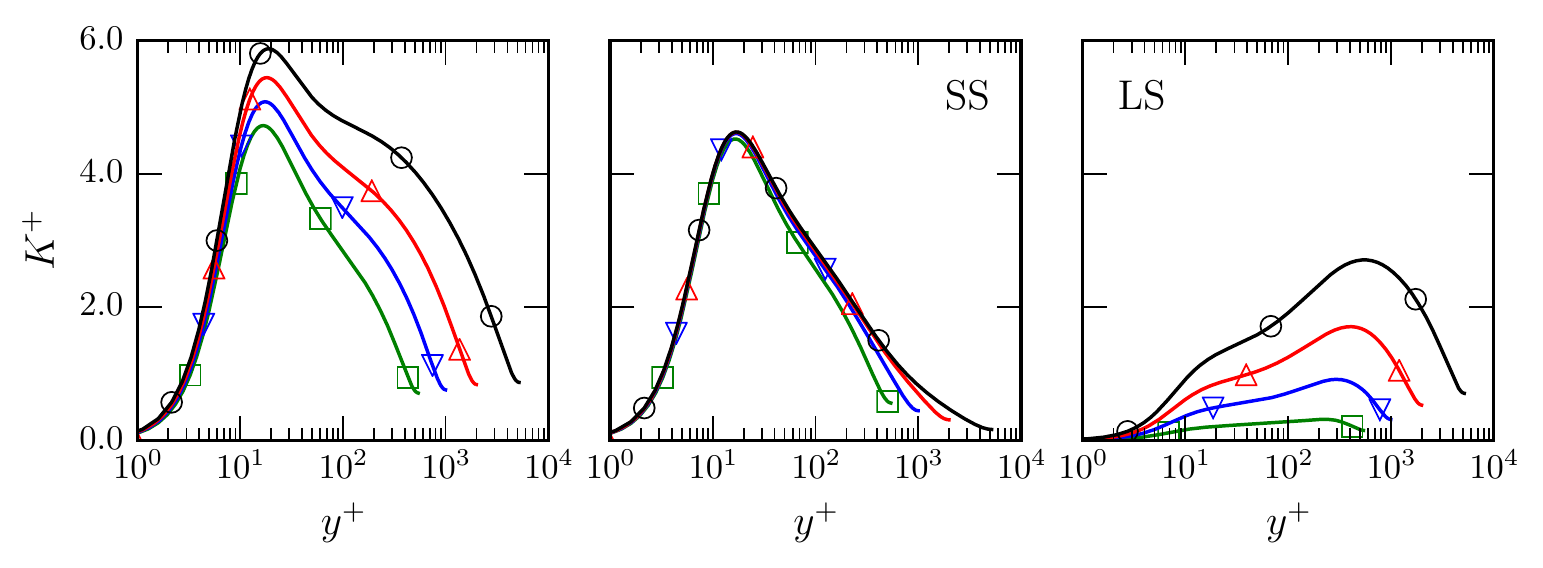}
  \end{center}
\caption{Unfiltered, high-pass (SS), and low pass (LS) portion of the 
turbulent kinetic energy scaled in viscous units 
with $k_{\text{cut}} \delta_{\nu} =0.00628$ ($\lambda_{\text{cut}}/\delta_{\nu} =1000$) 
for a channel flow at various $Re_{\tau}$ 
(green: $Re_{\tau}=550$, blue: $Re_{\tau} = 1000$,
red: $Re_{\tau}=2000$, black: $Re_{\tau}=5200$).
The contributions from the large scales increases
with $Re_{\tau}$, but the contribution from the small scales is largely 
independent of $Re_{\tau}$. Figure reproduced  
from \citet{Lee:2019} with permission.}
\label{fig:MK_filtered_TKE} 
\end{figure}

Thanks to advances in experimental techniques and computational power, 
the understanding of the physics of wall bounded flows has increased
greatly since the earliest investigations by \cite{Hagen:1839}, \citet{Darcy:1854}, and \citet{Reynolds:1895},
and the later work by \citet{Millikan:1938tv}. 
It is well known that there is an autonomous near-wall cycle of self sustaining
mechanisms \citep{Moin:1991, Hamilton:1995vu, Jeong:1997uj},  
involving low and high speed streamwise velocity 
streaks and coherent structures of quasi-streamwise vorticity. 
\citet{Jimenez:1999wf} showed that 
this cycle of near-wall dynamics persists without any input from 
the turbulence farther away from the wall. Moreover, if any element
of the cycle is suppressed, the near-wall turbulent kinetic energy (TKE) 
decays, and the flow becomes laminar. 
However, the large-scale structures (superstructures) in the
outer layer do modulate the turbulent fluctuations in the near-wall region
\citep{Hutchins:2007kd, Marusic:2010bb, Ganapathisubramani:2012dh},
leaving their ``footprint" on the autonomous
cycle. \citet{mathis_hutchins_marusic_2011} (see also references
therein) modulated a ``universal signal'' identified in experimental
data as the contribution of the small-scale near-wall turbulence, to
formulate a predictive statistical model.  The large-scale modulation
results, for instance, in an $Re_{\tau}$-dependent peak of the
turbulent kinetic energy in the near-wall region (see
figure~\ref{fig:MK_filtered_TKE}) because its influence increases with
increasing $Re_{\tau}$ \citep{DeGraaff:2000wm}.

Recently, \citet{Lee:2019} 
 performed spectral analysis of channel flow DNS data
for several different $Re_{\tau}$ (ranging from approximately 
$550$ to $5200$) to 
investigate the relative importance of different 
length scales to the production, transport, and dissipation 
of TKE.
Their results suggest that the small scales in the near-wall 
region behave universally. Indeed, when the energy spectrum 
is high-pass filtered
to only include contributions from wavenumbers with magnitude larger than some 
$k_{\rm cut}\delta_{\nu}= 0.00628$ (corresponding to a wavelength $\lambda_{\rm cut}/\delta_{\nu} = 1000$),
the resulting energy is found to be independent of \retau, as shown in figure~\ref{fig:MK_filtered_TKE}.
Similar results were also obtained in \citet{Samie:2018} for experimental 
data ranging from $Re_{\tau} \approx 6000-20000$. 
These works, along with those mentioned above, indicate that the near-wall region has universal
small scales,
independent of $Re_{\tau}$. The large scale
portion of the near-wall turbulence, however, is the result of eddies
whose size and influence on the turbulent statistics depend on
$Re_{\tau}$.
These observations suggest that an appropriately formulated numerical simulation
of only the small-scale near-wall dynamics should be able to reproduce
the near-wall small-scale statistics (e.g. SS in
figure~\ref{fig:MK_filtered_TKE}). One objective of the work reported
here is to test this hypothesis.

With this in mind, we endeavor to design a computational 
model of the universal near-wall small scales of turbulent, 
wall bounded shear flows. The 
primary modeling goal is to accurately represent the contribution of the 
small scales to the near-wall turbulent statistics without
simulating the entire wall-bounded turbulent shear flow.
The model is formulated to simulate wall bounded turbulence only in a 
near-wall, rectangular domain $\Omega$ localized to the boundary. The size of the 
domain scales in viscous units, so that as $Re_{\tau}$ increases, the 
domain shrinks in size relative to the size of the overall
flow whose near-wall turbulence is being modeled. 

The model is
formulated to mimic the mean flux of momentum from the outer layer, 
but it otherwise ``isolates'', or decouples, the near wall-wall dynamics from 
large-scale outer-layer influences, such as the modulations by superstructures. In
this way it is similar to the numerical experiments described
in \citet{Jimenez:1999wf} in which the equations of motion are
filtered to suppress the dynamics above some fixed wall-normal
height. It is fundamentally different from a low Reynolds number
channel flow, for example, whose dynamics are influenced by the
presence of the opposite wall.  If such a configuration can
accurately model the dynamics of the near-wall, small scale features
of the flow, it could be used to study the response of
near-wall turbulence to changes in the momentum environment,
including the effects of pressure gradients. Further, assuming a
separation of scales between the small-scale autonomous near-wall
dynamics and the large-scale outer turbulence that modulates it, the
model flow could be used to inform a representation of wall turbulence
in a wall-modeled large eddy simulation.

This paper reports on the development and evaluation of just such a
computational model of the near-wall layer in turbulent shear flows.
It originated as a design for the high-fidelity, ``microscale"
component of a multiscale computational approach for simulating wall
bounded turbulence in the style of the heterogeneous multiscale
method \citep{abdulle:2012}, as pursued
by \citet{sandham:2017}, who coupled their microscale model to a
large eddy simulation (LES) in a full channel.  Previous multiscale
approaches of this type include \citet{pascarelli:2000}
and \citet{tang_akhavan_2016}, in which large eddy simulations are
coupled to minimal flow unit simulations.  One application
of the model will be to generate data to inform a
pressure-gradient-dependent wall model for an LES
\citep{Piomelli:2002,Bose:2018}, as suggested
by \citet{Coleman:2015eo}. In this case, the model plays the role of
the universal signal in \citet{mathis_hutchins_marusic_2011}.
Additionally, the current model approach could be
used to study the interaction between the small, near-wall
turbulent dynamics and more complicated physical processes, such as
heat transfer, chemical reactions, turbophoresis, or surface
roughness.


The rest of the manuscript is organized 
as follows: section 2 contains a description
of the computational model and the numerical method used to integrate
the equations of motion. Section 3 provides a comparison between
the statistics generated by the model and the corresponding 
quantities from DNS for the cases of both zero and mild favorable pressure gradients. 
In section 4 the results are summarized, and possible applications
and extensions of the model are discussed. 


\section{Formulation}

\subsection{Notation}
In the following discussion, the velocity components in the
streamwise (x), wall-normal (y) and spanwise (z) directions
are denoted as $u$, $v$, and $w$, respectively, and when
using index notation, these directions are labeled 1, 2, and
3, respectively. The expected value is denoted with 
angle brackets (as in $\langle \cdot \rangle$), and upper
case $U$ and $P$ indicate the mean velocity and pressure, so 
that $\langle u_i \rangle = U_i$. The velocity and pressure 
fluctuations are indicated with primes, e.g. $u_i = U_i + 
u_i'$. Partial derivatives shortened to $\partial_i$ signify
$\partial/\partial x_i$, differentiation in the direction $x_i$.
 The mean advective derivative is
$D(\cdot)/Dt = \partial_t (\cdot) + U_j \partial_{j} (\cdot)$, where
Einstein summation notation is implied. In general, repeated indices
imply summation, with the exception of repeated Greek indices.
Lastly,
the superscript `$+$' denotes non-dimensionalisation with the 
kinematic viscosity $\nu$ and the friction velocity $u_{\tau}$. 

\subsection{Motivation}\label{sec:motivation}

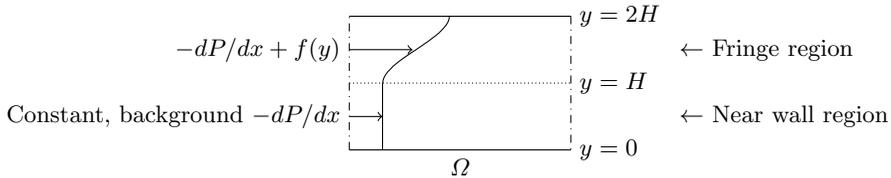
\begin{figure}
\centering
\begin{tikzpicture}[xscale=4.4,yscale=4.4]
\draw[dashdotted](0,0) -- (0,0.4);
\draw (0,0.4) -- (.666,0.4);
\draw[dashdotted] (.666,0.4) -- (.666,0.0);
\draw (0,0.0) -- (.666,0.0);
\draw[densely dotted] (0,0.2) -- (.666,0.2);
\draw[variable=\y, domain=0:0.2] plot({0.10},{\y});
\draw[variable=\y, domain=0.2:0.4] plot({-50*(\y*\y*\y - 0.9*\y*\y + 0.24*\y - 0.02) + .1},{\y});
\draw[->] (0,0.1) -- (0.1,0.1);
\draw[->] (0.,0.3) -- (0.19,0.3);
\draw (0., 0.1)node[left]{Constant, background $-dP/dx$} ;
\draw (0.966, 0.1)node[right]{$\leftarrow$ Near wall region} ;
\draw (0., 0.3)node[left]{$-dP/dx + f(y)$} ;
\draw (0.966, 0.3)node[right]{$\leftarrow$ Fringe region} ;
\draw (.666,0)node[right]{$y=0$};
\draw (.6660,0.2)node[right]{$y = H$};
\draw (.6660,0.4)node[right]{$y = 2H$};
\draw (.333, 0)node[below]{$\Omega$};
\end{tikzpicture}
\caption{
The fluid is subject to periodic boundary conditions at the 
(dash-dotted) side walls, constant Dirichlet/Neumann conditions
at the upper boundary $y=2H$, and the no-slip condition 
at the wall $y=0$, as described in (\ref{eq:eq_of_motion}).
In addition to the constant pressure gradient assumed to be present in the 
near wall layer, the model includes an auxiliary pressure 
gradient in a ``fringe region"
to account for momentum transport at the computational boundary $y=2H$.}
\label{fig:fringe_region} 
\end{figure}
Intrinsic to the computational model is the assumption of a
  separation of temporal and spatial scales between the small-scale
  turbulence arising from the autonomous near-wall dynamics and the
  large-scale outer-layer turbulence. The near-wall dynamics can
  therefore be considered to be in local equilibrium with the pressure
  gradient and momentum flux environment in which they are
  evolving. Furthermore, under this scale-separation assumption, both
  the local pressure gradient and the turbulent momentum flux from the
  outer layer toward the wall can be considered constants on the scale
  of the near-wall dynamics being simulated. Note that despite this
  assumption, the near-wall model will be representative of the wall
  layer in wall bounded flows that are not in equilibrium overall, or
  those with non-constant pressure gradients. The assumptions break
  down, for example, for a boundary layer near separation.

The goal of the computational model is to simulate the 
turbulent small scales in the near wall 
region as a function of an imposed pressure gradient only in a small, 
rectangular domain localized to the boundary. This 
necessarily means placing nonphysical computational boundaries 
in a region of chaotic, highly nonlinear dynamics. In addition to 
the standard no-slip condition at the lower boundary $y=0$, the use of periodic boundary
conditions at the side-walls is well established, assuming
the flow is statistically homogeneous in these directions. The problem of 
prescribing appropriate boundary conditions at the upper computational boundary,
however, is nontrivial \citep{Berselli:2006mathematics,Sagaut2006:multiscale}.
Once a mathematically well posed condition is prescribed, care must
be taken to prevent the approximation inherent in the boundary condition from polluting the turbulent dynamics 
in the domain's interior. 
To address this issue, the model augments the
near wall computational domain with a ``fringe region." 
In this fringe region, the flow 
is externally forced to account for the mean 
flux of momentum through
 the upper computational boundary that is precluded by the boundary conditions
imposed there.
The inclusion of such a region increases the computational cost
of the model, but it provides the  
momentum transport needed to
create the ``correct environment" for the evolution of turbulence in the near
wall region. In this way, the fringe region  
mollifies the effect of the nonphysical computational boundary. Similar 
techniques are used 
for designing inflow/outflow conditions in the DNS of 
turbulent boundary layers \citep{Oberlack:2004,Colonius:2004,Wu:2009fn,sillero:2013}, for example, as well as 
in molecular dynamics simulations, 
often referred to as a ``heat-bath" or ``thermostat"
\citep{Berendsen1984,thermostats_yong:2013}.

If one is interested in the turbulent statistics resulting from 
a constant pressure gradient near wall region out to a wall-normal 
height of $y \approx H$, 
the fringe region consists of 
a layer from $H \le y \le 2H$
in which 
a horizontally uniform streamwise forcing is applied,
as illustrated in figure \ref{fig:fringe_region}.

\subsection{Mathematical formulation}
The near wall patch (NWP) model is defined by the 
equations of motion and boundary conditions in the rectangular domain 
$\Omega = [0,L_x]\times[0,L_y]\times[0,L_z]$:
\begin{equation} \label{eq:eq_of_motion}
\begin{cases}
\begin{array}{rr}
\partial_t u_i + u_j \partial_j u_i + \partial_i p - \nu \partial_j \partial_j u_i  
=  f_i-\partial_i P & 
\text{in } \Omega  \\
\partial_i u_i = 0&  \text{in } \Omega \\
\partial_i P = dP/dx \,\, \delta_{i1} \text{ is constant}&
\text{in } \Omega  \\
u_i \text{ periodic in } x \text{ and } z \text{ directions}& 
\text{in } \Omega\\
u_i = 0& y = 0  \\
v = \partial_y w = 0&   y = L_y \\
\partial_y u = \psi \in \mathbb{R}&  y = L_y \\
f_i(x,y,z,t) = f(y) \, \delta_{i1}&  
\text{in } \Omega.
\end{array}
\end{cases}
\end{equation}

These are simply the forced incompressible Navier Stokes equations on a periodic domain 
in the wall-parallel directions, with the no-slip boundary condition 
at $y=0$ (the wall), and no-flow through and constant viscous tangential 
traction $\nu\, \psi$ in the streamwise direction ($x$) specified at the 
top $y=L_y$. The term $dP/dx$ models the externally imposed
  streamwise  pressure gradient
in the real turbulent flow being modeled, which is constant on the
scale of the NWP. The pressure $p$ is the NWP model's pressure field,
which is determined from the incompressibility constraint, in the
usual way.
It only remains to specify $\psi$ and the forcing function 
$f(y)$. The former represents the viscous flux of momentum through
the top, and the latter is a source of streamwise momentum 
that makes up for the missing turbulent flux of streamwise momentum 
through the upper computational boundary, owing to the boundary conditions that imply 
that the Reynolds stress vanishes there. 

The forcing function $f(y)$ is non-zero only in the fringe region
$y>L_y/2$, and is constructed such that
\begin{equation}\label{eq:f_constraint}
\int_0^{L_y}f(y)\,dy=-\tau_{\rm turb}, 
\end{equation}
where $\tau_{\rm turb}$ is the
turbulent flux of mean momentum through $y=L_y$ in the turbulent
flow being modeled.

In the real turbulent flow with 
a (locally) constant mean pressure gradient, the mean streamwise
momentum equation integrated over $[0,L_y]$ yields:
\begin{align}
  -\frac{d P}{d x}L_y + \left(\left.\nu\frac{\partial
    U}{\partial y}\right|_{y=L_y}-\tau_{\rm turb}\right) - \tau_w = 0& \\
\implies \tau_w = -\frac{d P}{d x}L_y + \tau_{\rm top}&, \label{eq:target_tau_w}
\end{align}
The term in parentheses, $\tau_{\rm top}$, is the total momentum flux
(viscous plus turbulent) through $y=L_y$, and it, along
with $dP/dx$, determines the mean wall shear stress $\tau_w$. 

The mean streamwise stress balance for the NWP model system \eqref{eq:eq_of_motion} is 
\begin{equation}\label{eq:model_mean_mom}
\nu \frac{\partial U}{\partial y} -\chevron{u' v'}
+ \int_{0}^{y} f(s) ds = y \frac{dP}{dx} + \tau_w.
\end{equation}
The boundary conditions in \eqref{eq:eq_of_motion} and the constraint
\eqref{eq:f_constraint} imply that at $y=L_y$, \eqref{eq:model_mean_mom} becomes
\begin{align}
\nu \psi + \int_{0}^{L_y} f(s) ds = L_y \frac{dP}{dx} + \tau_w& \label{eq:model_mean_mom_at_Ly} \\
\implies \tau_w = -\frac{d P}{d x}L_y + \nu \psi - \tau_{\rm turb}&, \label{eq:model_tau_w}
\end{align}
so that for fixed $L_y$, the parameters $\nu$, $\tau_{\rm turb}$, $\psi$, and $dP/dx$ 
determine $\tau_w$. Dimensional analysis therefore implies that there is
a two-parameter family of possible turbulent flows to model. 

By specifying $\psi$ and $\tau_{\rm turb}$ so that $\nu \psi - \tau_{\rm turb} = \tau_{\rm top}$, 
the NWP model's mean stress balance augmented with the forcing function $f$ \eqref{eq:model_tau_w} will 
be consistent with that of the real turbulent flow being modeled \eqref{eq:target_tau_w}. 

\subsection{Physical parameters}
The total mean stress at $y=L_y$ for the NWP model is given by 
\begin{equation}\label{eq:tau_tot_defn}
\tau_{\rm tot} := \nu \psi - \tau_{\rm turb}.
\end{equation}
In wall units, $\tau_{\rm tot}$
is simply a function of $dP^+/dx^+$:
\begin{equation}\label{eq:total_stress_specified}
\tau^+_{\rm tot} = 1 + L_y \frac{dP^+}{dx^+}. 
\end{equation}

The pressure gradient $dP^+/dx^+$ is thus one parameter defining a model case, and its values
for the four cases presented--
three favorable pressure gradient cases and one zero pressure gradient case--are 
shown in table \ref{table:simulation_cases}. The values for $\tau_{\rm tot}^+$ are 
also shown for convenience, but of course they are simply determined via 
\eqref{eq:total_stress_specified}.

The second parameter to define a NWP model case is $\psi^+$; 
it determines the portion of $\tau^+_{\rm tot}$ carried by the mean viscous 
stress. For all of the statistics reported in section \ref{sec:numerical_results}, 
however, the actual value of $\psi^+$ used was found to be insignificant. 
For each case listed in table \ref{table:simulation_cases}, 
an initial $\psi^+_{\rm dns}$ was determined from available
DNS data. 
Results were then compared from runs with 
 $\psi^+ = \psi^+_{\rm dns}$, 
$\psi^+ = 2\psi^+_{\rm dns}$, $\psi^+ = \psi^+_{\rm dns}/2$, and 
$\psi^+ = 0$ , and the differences
 were found to be negligible. Hence, there is really only a one-parameter
family of possible turbulent flows to explore with the model, and for simplicity, $\psi^+$ 
is set to zero. Accordingly, \eqref{eq:tau_tot_defn} reduces to 
\begin{equation}\label{eq:tau_tot_reduction}
 \tau_{\rm tot} = - \tau_{\rm turb}. 
\end{equation}

\begin{figure}
  \begin{center}
    \includegraphics[width=1.0\textwidth]{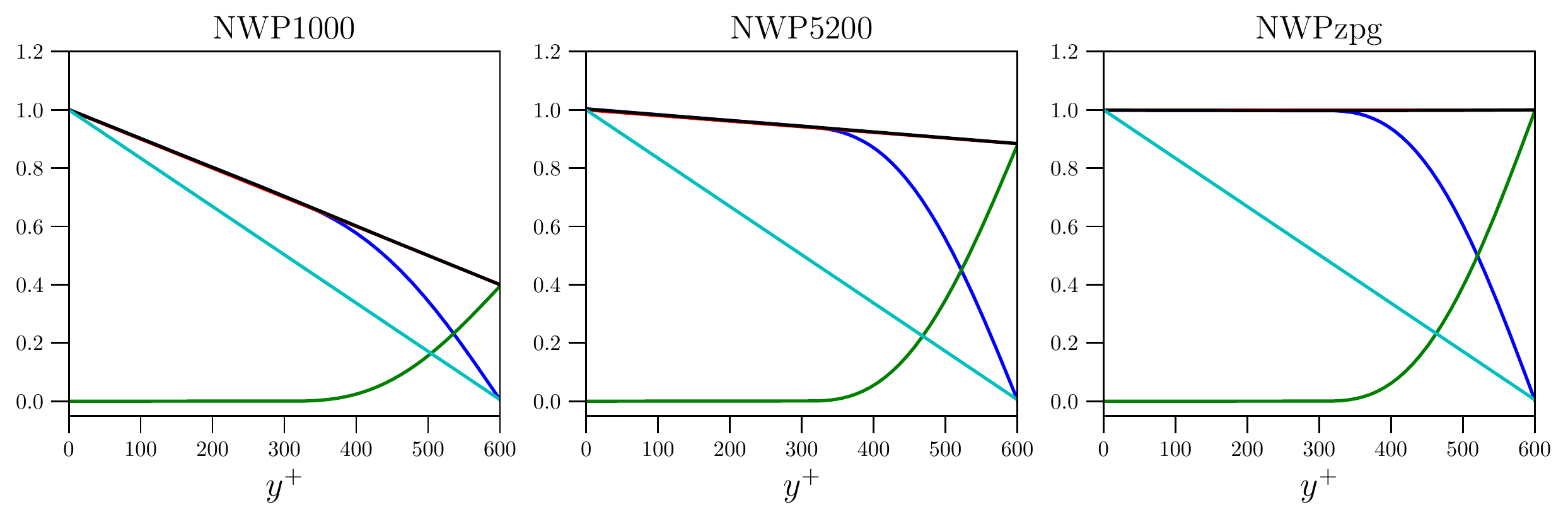}
  \end{center}
\caption{
Blue: model total stress $\tau^+_{\text{model}}(y^+)=\partial U^+/\partial y^+ - \chevron{u'v'}^+$; 
Red: target total stress $\tau_{\rm target}^+=1 + y^+\, dP^+/dx^+$;
Green: Primitive $F^+(y^+)$ of forcing function $f^+$;
Black: $\tau^+_{\text{model}}(y^+) + F^+(y^+)$;
Cyan: total stress for a simulation \textit{without} the forcing function $f^+$;
for all the three model cases listed in table
\ref{table:simulation_cases}.
}
\label{fig:stress_balance} 
\end{figure}

Figure \ref{fig:stress_balance} illustrates
the statistically converged stress balances for these model cases. Included in the figure is 
a stress profile resulting from simulating the equations of
motion \eqref{eq:eq_of_motion} \textit{without} the extra
momentum flux provided by the auxiliary pressure gradient $f$
 (the cyan curve). In 
that case, a similar analysis to that shown in equations \eqref{eq:model_mean_mom}--\eqref{eq:model_tau_w} above shows that the
 statistically converged stress profile is simply 
a linear function with values $\tau^+_w=1$ and $\psi^+$ at $y^+=0$ and $y^+=L_y^+$, 
respectively. It is clear that this stress profile is in poor agreement with the
target profiles \eqref{eq:total_stress_specified}, which is to be expected, 
since $\psi^+ = 1/(\kappa L^+_y)$ in the log-region. Moreover, 
this discrepancy increases with decreasing $dP^+/dx^+$ 
(corresponding to increasing $Re_{\tau}$ when comparing to a channel flow), illustrating
the utility of including the extra momentum flux provided by the
auxiliary forcing term $f$.

\begin{table}
  \begin{center}
  \def~{\hphantom{0}}
    \begin{tabular}{c c c c c}
   Case name         & NWP550  & NWP1000  & NWP5200 & NWPzpg             \\
\hline
   $-dP^+/dx^+$      & $(543.496)^{-1}$   &  $(1000.512)^{-1}$     & $(5185.89)^{-1}$ & 0~      \\
   $\tau_{\rm tot}^+$& -0.10396  & 0.4003        &  0.8843 &    1~  \\
    \end{tabular}
    \caption{Imposed pressure gradient and the resulting total momentum flux 
$\tau_{\rm tot}$ at $y=L_y$ for the model cases presented. The 
favorable pressure gradient parameters were selected to match the
pressure gradients in the channel flow cases 
at \url{https://turbulence.oden.utexas.edu/}. Note that data from the NWP550
case is only used in figure \ref{fig:uprime_Ruu}.}
  \label{table:simulation_cases}
  \end{center}
\end{table}

\subsection{Computational parameters}
The remaining model parameters, consistent for all simulation cases, 
are summarized in table \ref{table:simulation_parameters}.
The size of the rectangular domain $\Omega$
is taken to be $L_x^+ = L_z^+ = 1500$ and $L_y^+ = 600$; 
these values, while somewhat arbitrary,
were determined through a combination of numerical experimentation and the spectral analysis
of \citet{Lee:2019}. Their work suggests that the contributions to the turbulent kinetic 
energy from modes with wavelengths $\lambda^+ < 1000$
are universal and $Re_{\tau}$-independent in a region below a wall-normal distance of 
approximately $y^+ = 300$.
 Accordingly, $L_y^+ = 2\cdot 300=600$ is chosen 
to allow for a sufficiently large fringe-region to mollify 
the effect of the nonphysical computational boundary at $y=L_y$ (see figure \ref{fig:fringe_region}), 
and both $L_x^+$ and $L_z^+$ are taken to be at least 1000.  

For the domain size in the stream/spanwise directions, there generally is a balance
between computational cost and the accuracy of the model, as defined by a comparison
of the model's energy spectral density with that of large scale DNS. In particular, a
variety of domain sizes were tested, ranging from $L_x^+=L_z^+=1000$ to approximately $3500$.
Increasing $L_x$ and $L_z$ results in better agreement of the model's 
low-wavenumber, large-scale portion of the energy spectral density with the corresponding
portion computed from DNS. The high-wavenumber, small-scale portion of the model's energy density, 
however, was insensitive to the domain size, so long as $L_x$ and $L_z$ were not taken 
to be too close to $L_x^+=L_z^+=1000$. 
In particular, $L_x^+=L_z^+=1500$ was found to be the smallest domain size
capable of reproducing
the universal small-scales discussed in section \ref{section:universal_small_scales}.

Given some target turbulent flux of mean momentum $\tau_{\rm turb}$, 
the auxiliary forcing $f$ must satisfy \eqref{eq:f_constraint}, but it is
otherwise unconstrained. 
For the simulations reported here, 
$f$ is defined explicitly to be 
\begin{equation}
f(y) =
\begin{cases} 
\begin{array}{rr}
4\tau_{\rm turb}/L_y^4  \left(L_y-2y\right)^2 \left(5L_y-4y\right), &y \in [L_y/2,L_y] \\
0, &y \in [0,L_y/2],
\end{array}
\end{cases}
\end{equation}
which was chosen to satisfy $f(L_y/2) = f'(L_y/2) = f'(L_y) = 0$. 
In particular the constraint $f'(L_y/2)=0$ is important
so that the transition in forcing from the near-wall region to the fringe region is smooth. Other
functional forms of $f$, however, are of course possible. In particular a quadratic 
profile satisfying \eqref{eq:f_constraint} and $f(L_y/2)=f'(L_y/2)=0$ was tested, and no detectable 
changes in the statistics in the near-wall region $y^+\in[0,300]$ were found. 

\subsection{Numerical implementation and resolution}
The model (\ref{eq:eq_of_motion}) is solved numerically using the 
velocity-vorticity formulation due to \citet{Kim:1987ub}. The 
equations of motion are discretized with a Fourier-Galerkin 
method in the stream/spanwise directions and a seventh order B-spline
collocation method in the wall-normal direction \citep{Kwok:2001510,Botella:2003dn,Lee:2015er}. 
They are integrated
in time with a low-storage, third order Runge-Kutta method that 
treats diffusive and convective terms implicitly and explicitly, 
respectively \citep{Spalart:1991wu}.
The numerical resolution in both space and time is consistent with 
that of DNS. The number of Fourier modes, 
and hence the numerical resolution,
used in each simulation is listed in table \ref{table:simulation_parameters}, 
and can be compared with, for instance, table 1 in \citet{Lee:2015er}. 
In addition, the collocation
 point spacing in the wall-normal direction 
is similar to previous DNS studies; the total number of collocation
points $N_y$ is taken to be equal to the number of collocation points below
$y^+ = 600$ in \citet{Lee:2015er}. They are then distributed
in the near wall region according to the same (shifted and rescaled) stretching function.

The model is implemented with a modified version 
of the PoongBack DNS code \citep{Lee:2013kv,Lee:2014ta}, and 
the initial condition is taken from a restart file from a DNS run 
that is truncated to fit in $\Omega$ at the resolutions listed 
in table \ref{table:simulation_parameters} and modified to satisfy 
the boundary conditions \ref{eq:eq_of_motion}. 

\begin{table}
  \begin{center}
  \def~{\hphantom{0}}
    \begin{tabular}{c c c c c c c c c}
     $\psi^+$    &  $L_x^+= L_z^+$  & $L_y^+$  & $N_x$ & $N_z$ & $N_y$ & $ \Delta x^+$ & $\Delta z^+$ & $\Delta y^+_{w}$     \\
     0~  &  1500~           & 600~     & 120   & 256   & 192   &    12.5~     &  5.86~       &  0.002817~                 \\
    \end{tabular}
    \caption{Summary of simulation parameters consistent for all simulation cases; 
$\psi$ is the prescribed value for the Neumann boundary
condition in \eqref{eq:eq_of_motion}. 
$N_x$ and $N_z$ refer to the number of Fourier modes, while $N_y$ is
the number of B-spline collocation points.
$\Delta x = L_x/N_x$ and similarly for $\Delta z$. 
 $\Delta y_w$ is the collocation point spacing at the wall.}
  \label{table:simulation_parameters}
  \end{center}
\end{table}


\subsection{Statistical convergence}
\begin{figure}
  \begin{center}
    \includegraphics[width=1.0\textwidth]{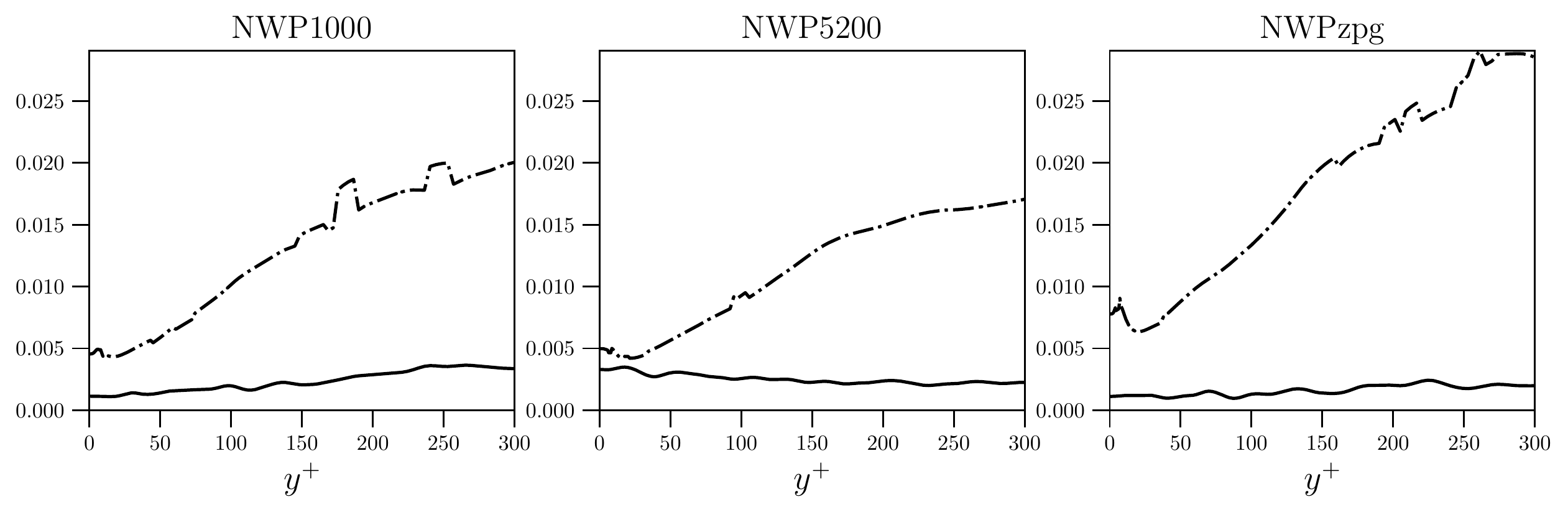}
  \end{center}
\caption{Statistical convergence for the model cases listed in 
table \ref{table:simulation_parameters}--
solid lines: Absolute error $|\tau^+_{\text{model}}(y^+) - \tau_{\rm target}(y^+)|$,
where $\tau^+_{\text{model}} = \partial U^+/\partial y^+ - \chevron{u'v'}^+$
and $\tau^+_{\rm target} = 1 + y^+ dP^+/dx^+$; 
dash-dotted lines: Standard
deviation of the estimated statistical error for $\tau^+_{\text{model}}$ 
in the region $y^+\in[0,300]$.
}
\label{fig:stress_error} 
\end{figure}

The method of \citet{Oliver:2014dh} is used to assess the uncertainty in the
statistics reported due to sampling noise. For each pressure gradient case, 
statistics are collected by averaging in time until the estimated statistical uncertainty
in the mean stress profiles is less than a few percent. For the cases 
in table \ref{table:simulation_parameters} reported here, the sampling error
is less than three percent, as shown 
in figure \ref{fig:stress_error}. 

\section{Numerical results}\label{sec:numerical_results}
The near-wall patch model can be interpreted in two ways. In the
first, one considers it to be a model for the small-scale near-wall
turbulence in a wall bounded flow with mean pressure gradient in wall
units the same as the imposed pressure gradient in the model. In this
case, one aspires to have the statistics from the model and the real
flow match for quantities that are insensitive to the unrepresented
large scales or for the statistics of the small scales computed from a
low-pass filter as in \citet{Lee:2019}. This is the interpretation
explored in the results reported here. In the second interpretation,
the NWP model represents small-scale near-wall turbulence in a region
of a real wall-bounded turbulent flow with local pressure gradient in
wall units based on the local wall shear stress the same as that
imposed in the model. In this case, the NWP model is analogous to the
universal signal of \citet{mathis_hutchins_marusic_2011}, representing
the process that is modulated by large-scale outer-layer fluctuations
in a real turbulent flow.  These interpretations are clearly
complementary, and both should be valid given the scale-separation
assumptions on which the model is predicated.

The statistics reported here were computed from the three near wall
patch model cases NWP1000, NWP5200, and NWPzpg. The first two
are so named because the imposed favorable pressure gradient in these
cases is the same in wall units as the mean pressure gradient in a fully developed
channel flow with friction Reynolds number $Re_\tau=1000$ and
5200. For NWPzpg, the imposed pressure gradient is zero. To assess the
quality of the NWP, statistics from the model will be compared to
those from the channel flow DNS of \citet{Lee:2015er,Lee:2019} at
$Re_{\tau}=1000$ and $Re_{\tau}=5200$--referred to below as LM1000 and
LM5200 (statistics at \url{https://turbulence.ices.utexas.edu}), and
the zero pressure turbulent boundary layer DNS
of \citet{sillero:2013,Borrell:2013ks,SIMENS20094218} at
$Re_\tau=2000$, which is referred to below as SJM2000 (statistics
at \url{https://torroja.dmt.upm.es/turbdata/blayers/high_re/}).
So, comparisons are being made to flows in which the mean
pressure gradient is the same in wall units as the imposed pressure
gradient in the NWP models. However, many wall shear flows can have
the same $\partial P^+/\partial x^+$. For example, the NWP1000 and
NWP5200 cases could equally well be associated with favorable pressure
gradient boundary layers and the NWPzpg case has imposed pressure
gradient consistent with an infinite Reynolds number channel. Note
that there is streamwise evolution of the mean wall shear stress in
SJM2000, which is not the case in the channel flow cases. The 
comparison here is hence predicated on this evolution occurring over scales
that are asymptotically large relative to the viscous scale, which
indeed is the case.

%


\subsection{Mean velocity and shear stresses}\label{section:mean_vel_shear_stress}
\begin{figure}
\begin{center}
\includegraphics[width=1.0\textwidth]{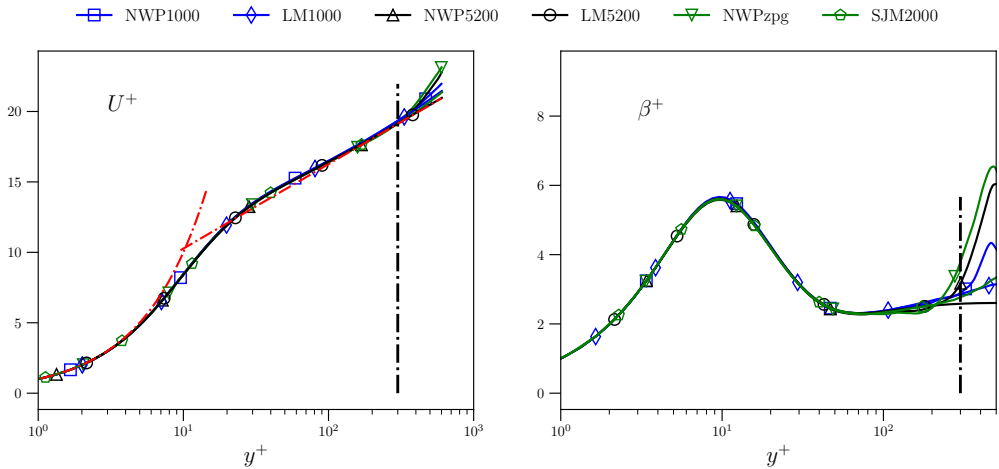}
\caption{Mean velocity $U^+$ (left) and the indicator function 
 $\beta^+ = y^+ \, \partial_{y^+} U^+$ (right) versus $\log(y^+)$. The black 
dashed-dotted vertical line marks $y^+=300$, and the red 
dashed-dotted lines plot the law-of-the-wall $U^+=y^+$ and 
$U^+ = (1/\kappa)\log(y^+) + B$, where $\kappa = 0.384$ and $B = 4.27$
\citep{Lee:2015er}.}
\label{fig:mean_U_and_indicator}
\end{center}
\end{figure}

\begin{figure}
\begin{center}
\includegraphics[width=1.0\textwidth]{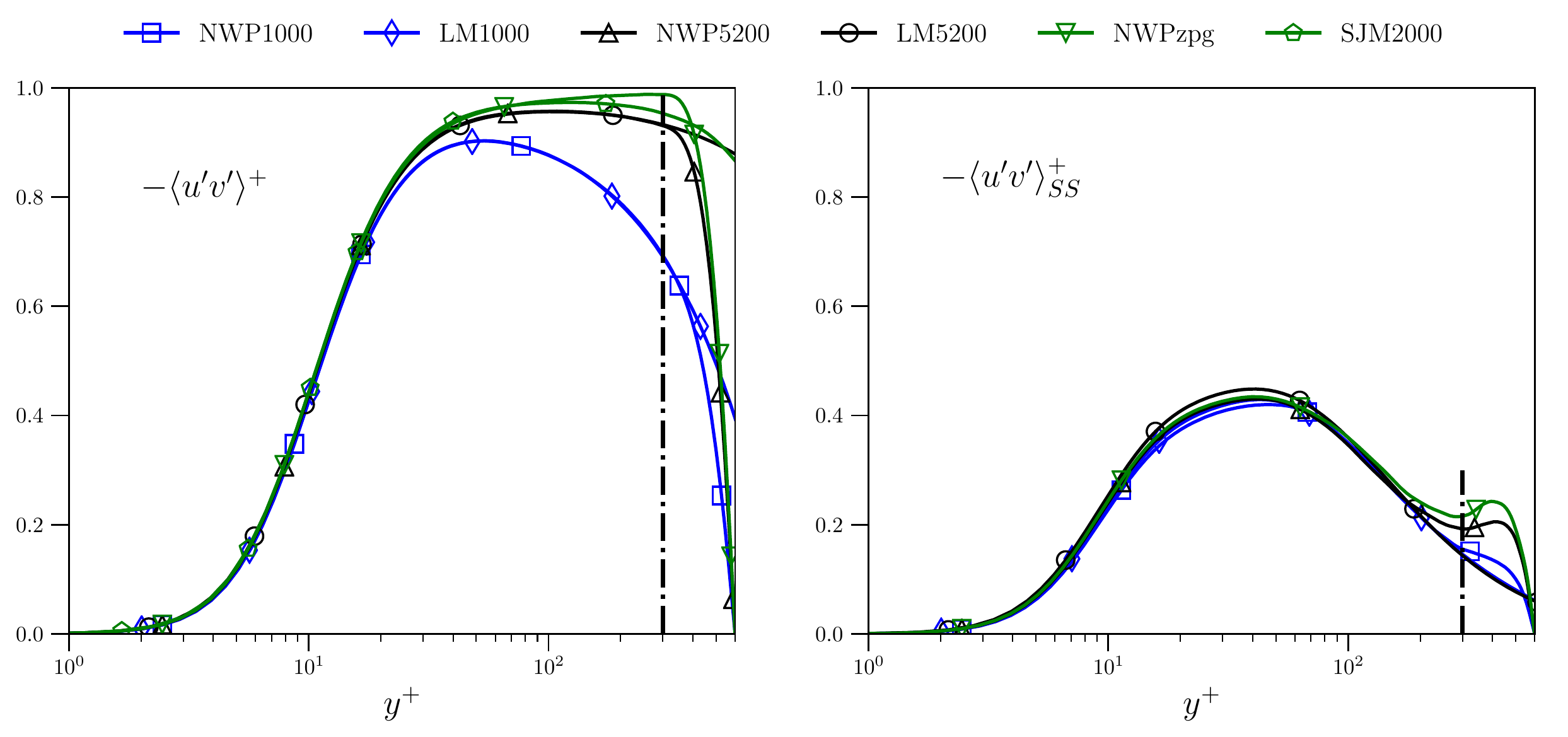}
\caption{(a) Reynolds stress $-\chevron{u'v'}^+$ and (b) the filtered
Reynolds stress $-\chevron{u'v'}^+_{SS}$ (defined
by \eqref{eq:uv_ss_defn} and discussed in section \ref{section:universal_small_scales}) 
as a function of $\log(y^+)$. The black dashed-dotted
vertical line marks $y^+=300$.
}
\label{fig:uv_comparison}
\end{center}
\end{figure}

If the near-wall turbulence fluctuations represented in the NWP
model dominate the Reynolds stress, as is expected from the spectral
analysis of \citet{Lee:2019}, then the mean velocity in the NWP
should match that in a full turbulent flow.
Figure \ref{fig:mean_U_and_indicator}
demonstrates that this is indeed the case.
The relative
error in $U^+$ is less than $0.6\%$ for $y^+ \in [0,300]$, and the error is similarly
small for the log-law indicator function $\beta^+$
\begin{equation}
\beta^+(y^+) := y^+\, \frac{\partial U^+}{\partial y^+} 
\end{equation}
in the range $y^+\in[0,100]$. However, there is
mild disagreement of $\beta$ in the range $y^+ \in [100,300]$. As expected, the profiles diverge
for $y^+>300$. 
The NWP model's Reynolds shear
stress $\chevron{u'v'}$ is in excellent agreement with that of the DNS
in the region $y^+\in[0,300]$, as expected given the agreement of the mean velocity; 
see figure \ref{fig:uv_comparison}. 
For the channel cases, the error
is less than $0.5\%$, and for the zero pressure gradient case the error is below $4\%$. 
In the former, the total stress at $y=L_y$ is known analytically as 
$\tau^+_{\rm tot} = 1 + L_y^+ dP^+/dx^+$ and is
used to define $f(y)$ in \eqref{eq:f_constraint}, \eqref{eq:tau_tot_defn}, and 
\eqref{eq:total_stress_specified}.
In the latter, the same relations are used, though they are approximate due to the streamwise
growth of the near wall layer in a boundary layer. This may explain 
the relatively larger discrepancy between the $\chevron{u'v'}$ profiles for the boundary layer.
In both instances, recall that $\chevron{u'v'}$ necessarily vanishes
at the upper computational boundary 
as a consequence of the boundary condition $v=0$ in \eqref{eq:eq_of_motion}. The accuracy 
of the Reynolds shear stress profiles in spite of this condition demonstrates
the utility of the forcing function $f$ in enabling momentum transport
to the near wall region $y^+\in[0,300]$.

\subsection{Energy spectral density}\label{section:energy_spectral_density}
For two points $(x,y,z),(x',y,z') \in \mathbb{R}\times [0,2\delta]\times \mathbb{R}$ in an infinitely
long channel, define the separation
distances $r_x = x-x'$ and $r_z = z-z'$. 
For a turbulent flow that is statistically homogeneous
in the stream and spanwise directions, the two point correlation
tensor
\begin{equation}\label{eq:two_pnt_correlation}
R_{ij}(r_x,y,r_z) := \chevron{u_i'(x+r_x,y,z+r_z) u_j'(x,y,z)}
\end{equation}
is a function only of $r_x$, $y$, and $r_z$. Taking the Fourier
transform of \eqref{eq:two_pnt_correlation} in the variables $r_x$ 
and $r_z$  defines the spectral
density $E_{ij}(k_x, y, k_z)$, which encodes the average contribution 
to the Reynolds stress tensor from different length scales
as a function of the wall-normal variable $y$.
The Reynolds stress tensor can 
be recovered by taking the limit $(r_x,r_z) \to (0,0)$ in 
\eqref{eq:two_pnt_correlation}, or by integrating the 
spectral density over all wavenumbers
\begin{equation}
\chevron{u_i'u_j'}(y) = \int\int E_{ij}(k_x,y,k_z) dk_x \, dk_z .
\end{equation}

For a wall bounded flow in a full size domain, 
the low-wavenumber contributions to the Reynolds stress 
represent the mean influences of the large scale structures on the near wall
dynamics. As is well known 
\citep{Hutchins:2007kd,Marusic:2010bb,LargescaleMotions:2017wl,Samie:2018,Lee:2019},
these low-wavenumber features of the near wall 
flow depend on $Re_{\tau}$. 
In contrast, there is
evidence that the high-wavenumber (small-scale) contributions 
to the Reynolds stress profiles are universal and independent
of $Re_{\tau}$ \citep{Samie:2018,Lee:2019}. 
 By design, the NWP model's domain size does not allow
accurate representation of the very large-scale structures known
to exist in the near wall region, and thus 
one cannot expect to capture their
influence on the near-wall velocity fluctuations.
 Instead, one expects the NWP model to correctly capture the dynamics
of the universal small scales  
elucidated by \citet{Samie:2018} and \citet{Lee:2019} associated with the autonomous cycle of 
\citet{Jimenez:1999wf}. 

To determine whether or not this is the case, the model's spectra
$E_{ij}$ are compared to their DNS counterparts. The spectra are
visualized in so-called log-polar coordinates \citep{Lee:2019}, in which the wavenumber
magnitude $k = \sqrt{k_x^2 + k_z^2}$ is represented on a logarithmic
scale. For fixed wall-normal location, the log-polar coordinates
are defined as 
\begin{align}\label{eq:log_polar_defn}
k_x^{\#} := \frac{k_x}{k} \log_{10}\left(\frac{k}{k_{\text{ref}}}\right) \\
k_z^{\#} := \frac{k_z}{k} \log_{10}\left(\frac{k}{k_{\text{ref}}}\right) \nonumber
\end{align}
where $k_{\text{ref}}$ is an arbitrary reference wavenumber that must be
smaller than the smallest nonzero wavenumber included in the spectrum, 
taken here to be $k^+_{\text{ref}}=1/50\,000$. Two advantages of these coordinates
are that lines of constant $k_z/k_x$ have slopes of $k_z/k_x$, and lines
of constant magnitude $k$ map to circles. In this way, the orientation and 
alignments of the Fourier modes are easily interpreted; see \citet{Lee:2019}
for a more detailed discussion. 

\begin{figure}
\begin{center}
\includegraphics[width=1\textwidth]{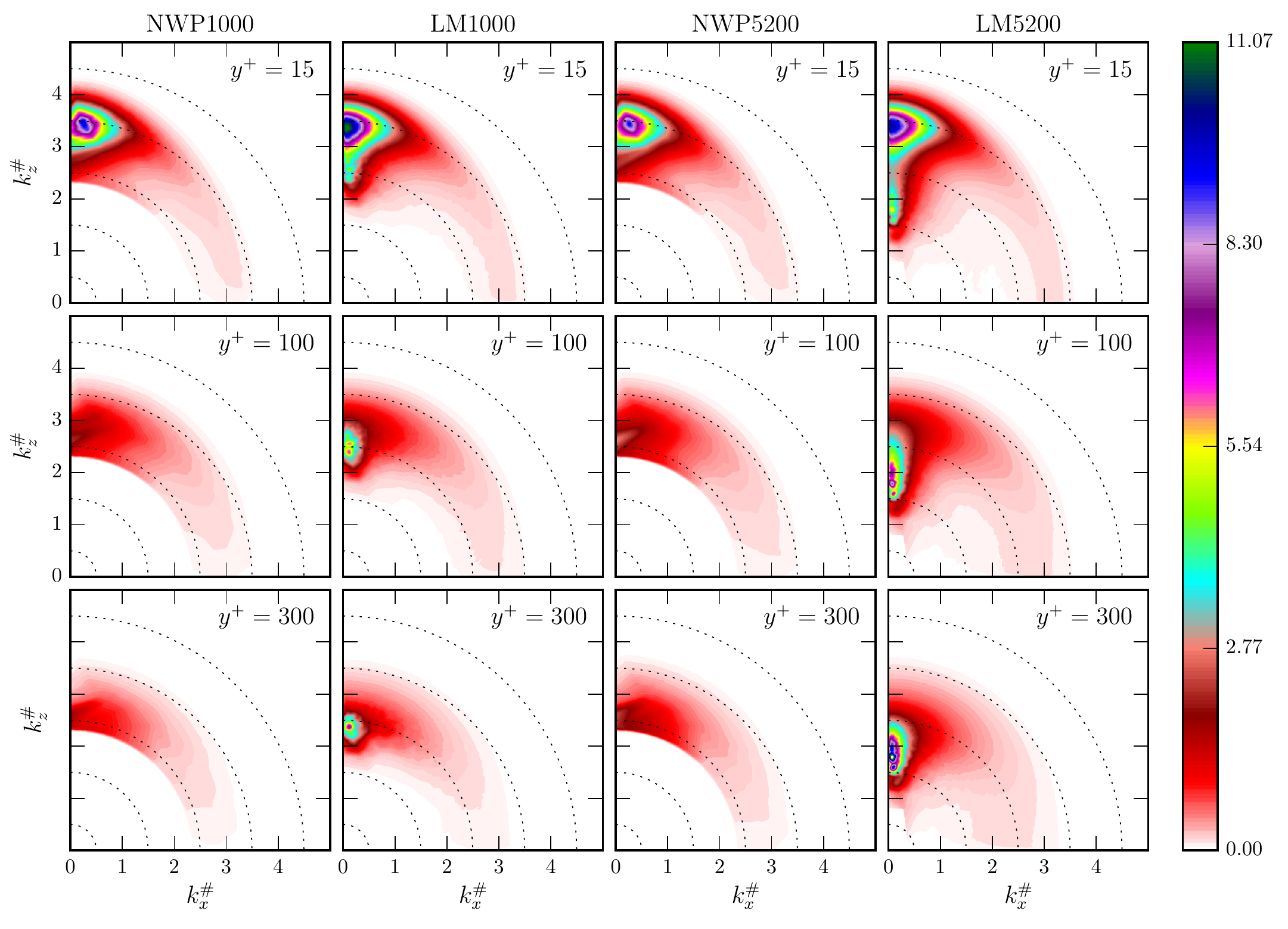}
\caption{Two-dimensional spectra of the streamwise velocity 
variance $\chevron{u'u'}^+$ in log-polar
coordinates, as defined by equation \eqref{eq:log_polar_defn}. 
$\lambda^+=10$ on the outermost dotted circle and increases
by a factor of 10 for each dotted circle moving inward, 
where $\lambda = 2\pi/k$ is the wavelength.}
\label{fig:2d_uu_wufl}
\end{center}
\end{figure}
\begin{figure}
\begin{center}
\includegraphics[width=1\textwidth]{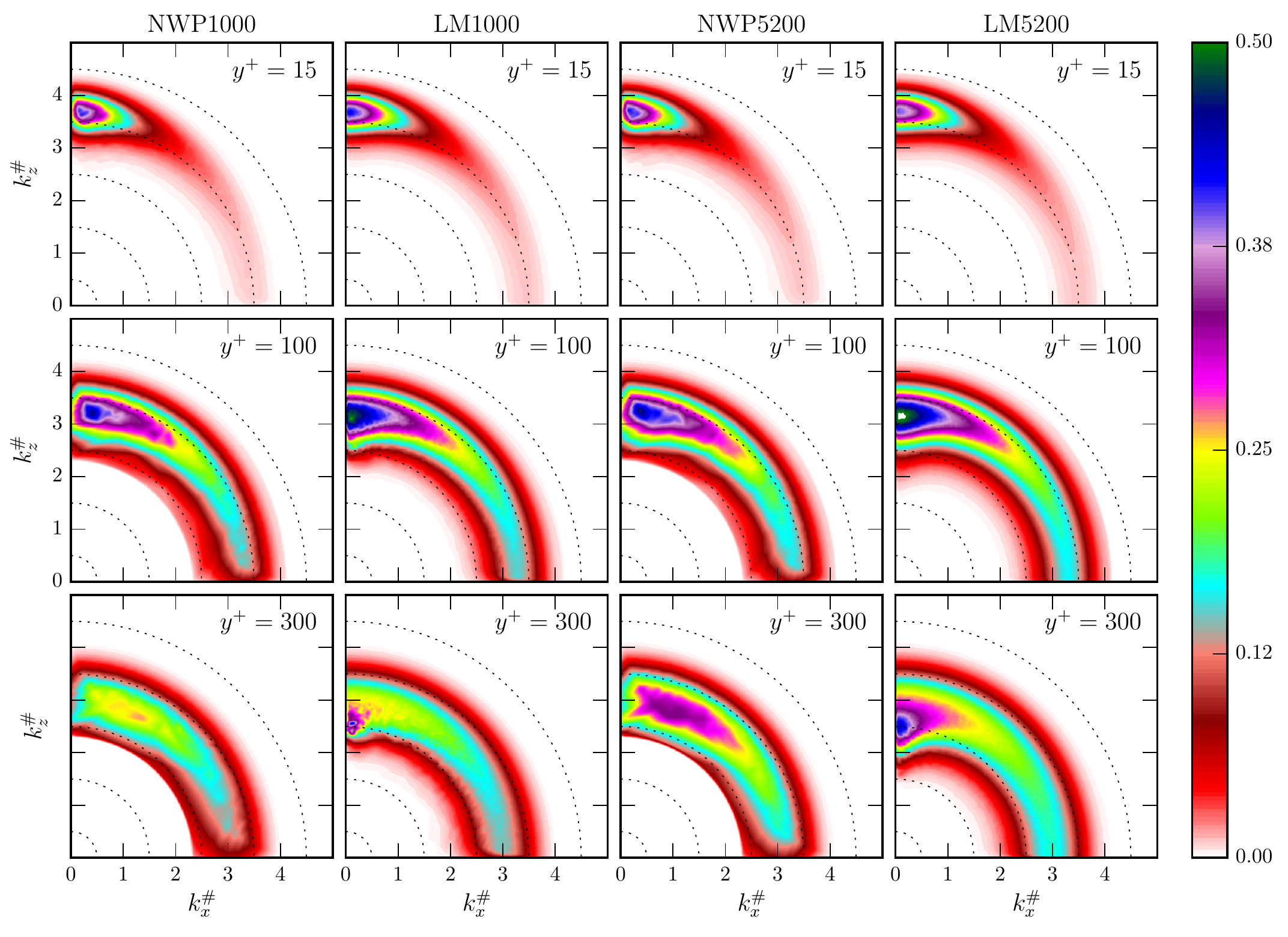}
\caption{Two-dimensional spectra of the wall-normal velocity
variance $\chevron{v'v'}^+$ in log-polar
coordinates, as defined by equation \eqref{eq:log_polar_defn}. 
$\lambda^+=10$ on the outermost dotted circle and increases
by a factor of 10 for each dotted circle moving inward,
where $\lambda = 2\pi/k$ is the wavelength.}
\label{fig:2d_vv_wufl}
\end{center}
\end{figure}
The two-dimensional spectral densities of the streamwise and wall-normal 
 velocity variances are shown in 
figures \ref{fig:2d_uu_wufl} and \ref{fig:2d_vv_wufl}, respectively. 
The spectra are visualized at the wall-normal 
locations $y^+ = 15$, $y^+=100$ and $y^+=300$ for the 
simulations NWP1000, NWP5200, LM1000, and LM5200. 

In each of the cases, the streamwise velocity spectra $E_{11}$ 
consist primarily of energy concentrated along
the $k_z^{\#}$ axis, with Fourier modes for which 
$k_z/k_x \gtrsim 10$ \citep{Lee:2019}. These correspond to structures 
that are strongly elongated in the streamwise $x$-direction,
such as the well-known, near wall low and high speed streaks. 
The channel flow data LM1000 and LM5200 (columns two and four
in figure \ref{fig:2d_uu_wufl}) show that this energy
exhibits two distinct features.
The first is an ``inner
energy site" \citep{Samie:2018}, a triangular shaped region in the near wall layer
$y^+\approx 15$  distributed primarily 
between wavelengths $\lambda^+ = 100$ and $\lambda^+ = 1000$ that
can be attributed to the autonomous
near wall dynamics described by \citet{Hamilton:1995vu,Jeong:1997uj,Jimenez:1999wf},
and others. The model $E_{11}$ spectra, 
shown in columns one and three in figure \ref{fig:2d_uu_wufl}, 
qualitatively reproduce the inner energy site, suggesting
that it captures the dynamics of the near wall, 
small scale energetic motions. 

The second feature is a concentration of energy at
relatively low wavenumbers (in the range $1000 < \lambda^+ < 10\,000$)
along the $k_z^{\#}$ axis at each of the wall-normal 
locations $y^+=15$, $y^+=100$, and $y^+=300$. 
These are due to the very large scale motions (VLSMs) imposed from 
the outer layer flow described by \citet{Hutchins:2007kd}
and \citet{Marusic:2010bb}. These VLSMs contribute
energy in the near wall region around $y^+=15$, and farther
away from the wall they are responsible for the majority 
of the turbulent kinetic energy. As both $y^+$
and $Re_{\tau}$ increase, the energy becomes more concentrated
 and is found at larger wavelengths, consistent with 
the attached eddy hypothesis of \citet{Townsend:1976uj}.
In addition, the VLSMs modulate
the near wall cycle through nonlinear interactions,
 creating large scale variations in the 
local wall shear stress that result in local variations in 
the dominant (most-energetic) wavelength \citep{Lee:2019}. 
Consequently, the spectral peak of the inner energy site 
for the DNS data is reduced and ``smeared out" 
as a function of $Re_{\tau}$. For example, the $Re_{\tau}=5200$ peak 
is about ten percent lower than the $Re_{\tau}=1000$ peak.

By design, the NWP model can only support modes with 
wavelengths less than or equal to $L_x^+ = L_z^+ = 1500$, 
meaning the VLSMs present in real wall bounded turbulence are not represented by the model. 
As a result, there is no energy associated with such
large scale structures;  
the concentration of energy at low wavenumbers along
the $k_z^{\#}$ axis (corresponding to wavelengths 
$\lambda^+\gtrsim 1000$) present in the DNS spectra is not present 
in that of the model. This is true both in the near wall region
and farther away from the wall at $y^+=100$ and $y^+=300$. 

Furthermore, the NWP model does not capture the nonlinear modulations of  
the autonomous cycle by the VLSMs. For instance, 
even though the model represents wavenumbers along the 
$\lambda^+=1000$ band, its spectra is not simply a spectral
truncation of the DNS spectra. Additionally, the peak of the 
inner energy site is nearly identical for the two model cases, 
differing by only a few percent. 
These differences between spectra of the model and DNS highlight
the important role the VLSMs play in the turbulent near wall layer.

The $E_{22}$ spectra 
are largest in the wavenumber regions in
which the $E_{11}$ spectra are peaked, as discussed in \citet{Lee:2019}.
Additionally,
the distribution of energy generally becomes more isotropic with 
increasing wall-normal distance $y^+$. 
Figure \ref{fig:2d_vv_wufl} shows that the DNS energy 
density $E_{22}$ is primarily, but not exclusively, located at the small
scales, i.\,e.\ at wavenumbers $\lambda^+<1000$. 
Because the NWP model adequately resolves such structures, its 
energy density $E_{22}$ is in overall good agreement
with the DNS spectra, especially in the near
wall region. Farther away from the wall the agreement is not 
as good since the DNS spectra are peaked at lower wavenumbers. 
Accordingly, the model's unfiltered $\chevron{v'v'}$ profiles shown in 
figure \ref{fig:vel_vars_filtered} show excellent agreement with the
corresponding DNS profiles in region $y^+ \in[0,300]$; they are
nearly identical for $y^+ \lesssim 50$ and only display slight 
discrepancies for $y^+ \in [50,300]$.

Lastly, fidelity of the NWP model's energy density $E_{33}$ (not shown)
 in reproducing the DNS spectra is similar to
the $E_{11}$ spectra. It clearly 
approximates the small scales in the near wall region well, 
but it fails to capture the 
modulation by the large scale structures at each wall-normal location. 

\subsection{Universality of small scales}
\label{section:universal_small_scales}

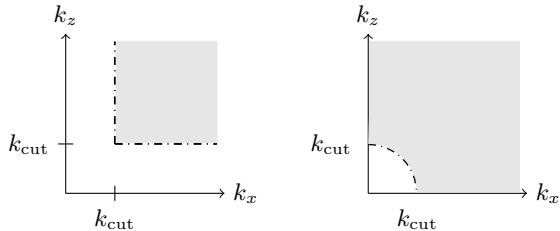
\begin{figure}
\centering
\begin{tikzpicture}
\draw[->] (0,0) -- (2.1,0)node[right]{$k_x$}; 
\draw[->] (0,0) -- (0,2.1)node[above]{$k_z$};
\draw[dashdotted, line width = 0.24mm] (0.65,0.65) -- (0.65,2); 
\draw[dashdotted, line width = 0.24mm] (0.65,0.65) -- (2,0.65); 
\begin{scope}
\filldraw [gray!20] (0.67,0.67) rectangle (2,2); 
\end{scope}
\draw (-0.1,0.65)node[left]{$k_{\rm cut}$} -- (0.1,0.65); 
\draw (0.65,-0.1)node[below]{$k_{\rm cut}$} -- (0.65,0.1); 

\draw[->] (4,0) -- (6.1,0)node[right]{$k_x$}; 
\draw[->] (4,0) -- (4,2.1)node[above]{$k_z$};

\coordinate (A) at (4, 0.65); 
\coordinate (B) at (4.65, 0.0); 
\draw[dashdotted, line width = 0.34mm, variable=\x, domain=4.0:4.65, name path=A] 
plot(\x, {sqrt(0.4225-(\x-4.)*(\x-4.))});
\draw[white, variable=\x, domain=4.01:4.65, name path=B] plot(\x, {1.0});
\tikzfillbetween[of=A and B]{gray!20}

\begin{scope}
\filldraw [gray!20] (4.02,1.00) rectangle (6,2.0); 
\end{scope}
\begin{scope}
\filldraw [gray!20] (4.65,0.02) rectangle (6,1); 
\end{scope}

\draw (3.9,0.65)node[left]{$k_{\rm cut}$} ; 
\draw (4.65,-0.1)node[below]{$k_{\rm cut}$} ; 
\end{tikzpicture}
\caption{
(Left) The shaded gray region indicates the subset (in the first quadrant) of 
wavespace $\mathcal{K}_{SS}$ defined 
by \eqref{eq:small_scale_wavemodes}.
(Right) the corresponding region defined instead by the circular, 
$L^2$ filter \eqref{eq:circular_filter} used in \citet{Lee:2019}.
}

\label{fig:K_ss} 
\end{figure}

To better quantify the universality of the 
small scales and assess the NWP model's ability 
to reproduce them, the energy
spectral density is high-pass filtered and then integrated
to measure the energy 
residing in 
the small scales. 
Let $\mathcal{K}$ denote the set of wavenumbers supported by a simulation, 
and let $k_{\rm cut} = 2\pi/\lambda_{\rm cut}$ with $\lambda^+_{\rm cut} = 1000$.
Define $\mathcal{K}_{SS}$ to be the subset of $\mathcal{K}$ with the property 
that $(k_x,k_z) \in \mathcal{K}_{SS}$ if 
\begin{equation}\label{eq:small_scale_wavemodes}
 \min\{|k_x|,|k_z|\} >k_{\rm cut},
\end{equation} 
visualized in figure \ref{fig:K_ss} (left).
The $\mathcal{K}_{SS}$ sets are meant to contain large wavenumbers associated with the universal 
small scales. Here $k_{\rm cut}$ is chosen based on the two-dimensional
spectra in \citet{Lee:2019}, where it is observed that 
the energy associated with the autonomous cycle has $\lambda^+ < 1000$. 

Note the high-pass filter \eqref{eq:small_scale_wavemodes}
is slightly different than the $L^2$ filter 
\begin{equation}\label{eq:circular_filter}
\sqrt{k_x^2 + k_z^2} > k_{\rm cut} 
\end{equation} 
used in \citet{Lee:2019} and visualized in figure \ref{fig:K_ss} (right).  
In particular, the wavenumbers on the axis $k_x=0$ (respectively $k_z=0$) 
with $k_z > k_{\rm cut}$ (resp. $k_x > k_{\rm cut}$)
are filtered out by \eqref{eq:small_scale_wavemodes} but not by \eqref{eq:circular_filter}. 
These axes contain the NWP model's approximation to the large scale motions present in a DNS
that do not ``fit" in the near wall patch domain. Such motions correspond to wavenumbers
smaller than $2\pi/L_x = 2\pi/L_z$, and they are 
not well represented by the NWP model. 
Hence, they are filtered out by \eqref{eq:small_scale_wavemodes}.
The approximation can be improved by increasing $L_x$ and $L_z$ (confirmed by 
numerical tests), although this of course increases the model's
overall computational cost.

Given $\mathcal{K}$ and $\mathcal{K}_{SS}$, the Reynolds stresses are
\begin{equation}
\chevron{u_i'u_j'}(y) = \sum_{(k_x,k_z)\in\mathcal{K}} E_{ij}(k_x,y,k_z),
\end{equation}
and the small scale energy can be quantified as 
\begin{equation}
\label{eq:uv_ss_defn}
\chevron{u_i'u_j'}_{SS}(y) = \sum_{(k_x,k_z)\in\mathcal{K}_{SS}} E_{ij}(k_x,y,k_z). 
\end{equation}

\begin{figure}
\begin{center}
\includegraphics[width=1\textwidth]{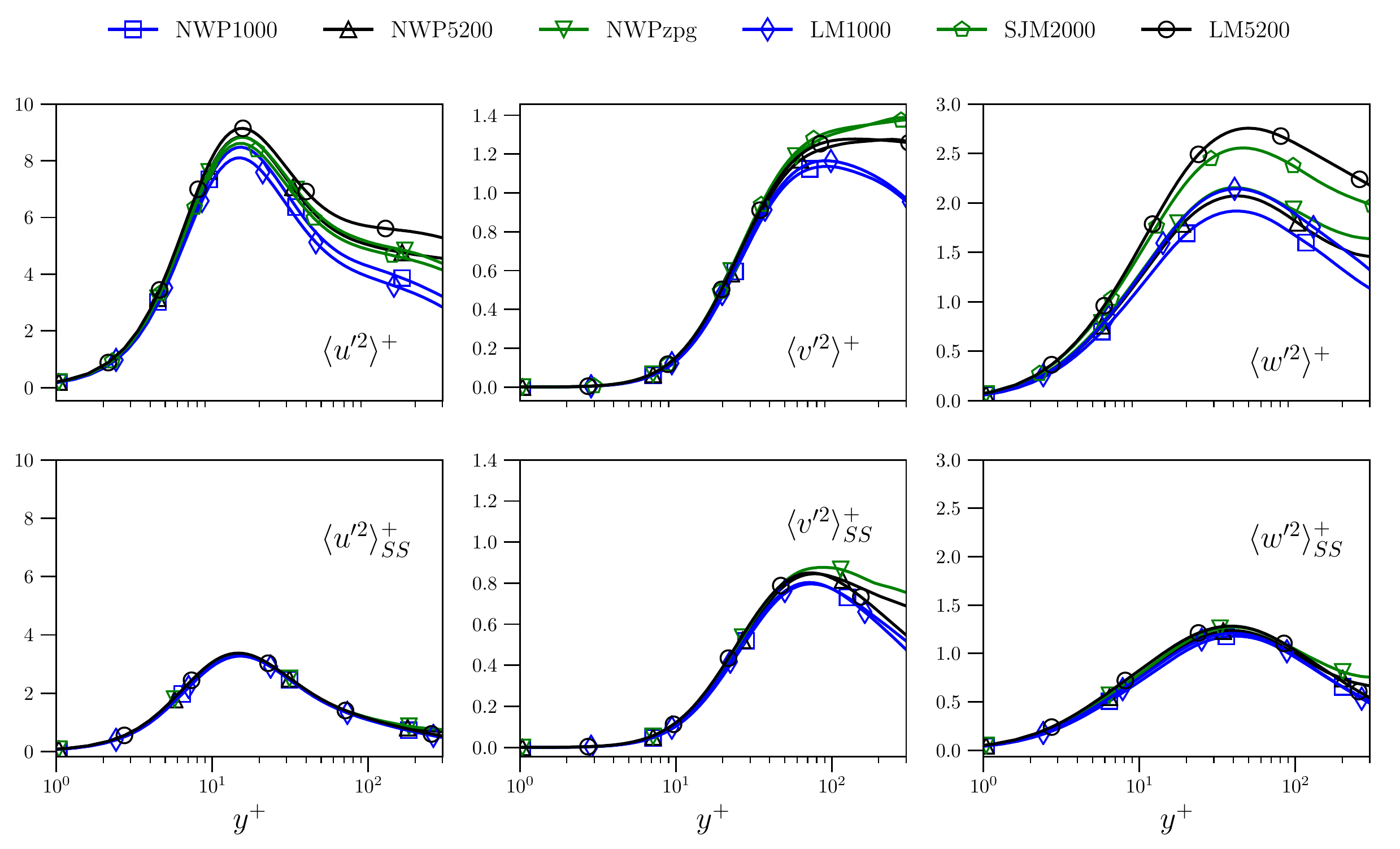}
\caption{Velocity variances $\chevron{u'_{\alpha}u'_{\alpha}}^+$ (top) 
and the corresponding high-pass filtered quantities $\chevron{u'_{\alpha}u'_{\alpha}}^+_{SS}$ 
(bottom) as a function of $\log(y^+)$.}
\label{fig:vel_vars_filtered}
\end{center}
\end{figure}
The velocity covariance $\chevron{u'v'}$ and variances 
$\chevron{u'_{\alpha}u'_{\alpha}}$, $\alpha=1,2,3$ and their
 high-pass filtered counterparts are shown in figures \ref{fig:uv_comparison}
and \ref{fig:vel_vars_filtered}, respectively. As previously mentioned, 
the model's
\textit{unfiltered} $\chevron{u'v'}$ and $\chevron{v'v'}$ profiles both agree
quite well with the corresponding DNS profiles.  
The model's streamwise and spanwise velocity variances, however, display nontrival 
discrepancies with the DNS profiles, as expected from the observed differences 
in the two dimensional spectra. In contrast, the model's high-pass filtered 
profiles all show agreement with the high-pass filtered DNS quantities. 
In all cases the agreement is excellent in the region $y^+\in[0,100]$, although
there are some relatively mild discrepancies for $y^+\in[100,300]$.
Moreover, it is clear that the high-pass filtered quantities are nearly 
$Re_{\tau}$ independent; the collapse of the $\chevron{u'u'}_{SS}$ profiles
is particularly convincing. Although two-dimensional spectra data is not
available for the zero-pressure gradient DNS case SJM2000, 
the $\chevron{u_i'u_j'}_{SS}$ profiles
computed from the model simulation NWPzpg are included for completeness; they display
the same universal behavior as the favorable pressure gradient flows. 
These observations lend support to the conclusion 
that the small scales in the near wall region are universal, and that
the difference in the Reynolds stress profiles as a function of $Re_{\tau}$
is due to the increasing influence of the VLSMs.
Previous results of this type obtained in both \citet{Lee:2019} and \citet{Samie:2018} 
involve high-pass filtering the \textit{entire} turbulent flow field, in which  
there are nonlinear interactions between wavenumbers across all the scales of motion. 
It is particularly
remarkable, however, that the NWP model reproduces the universal behavior of the small
scales \textit{without} the dynamic modulation of the near wall
autonomous cycle by the large scale structures.  


\subsection{Reynolds Stress Transport}
\begin{figure}
\begin{center}
\includegraphics[width=1\textwidth]{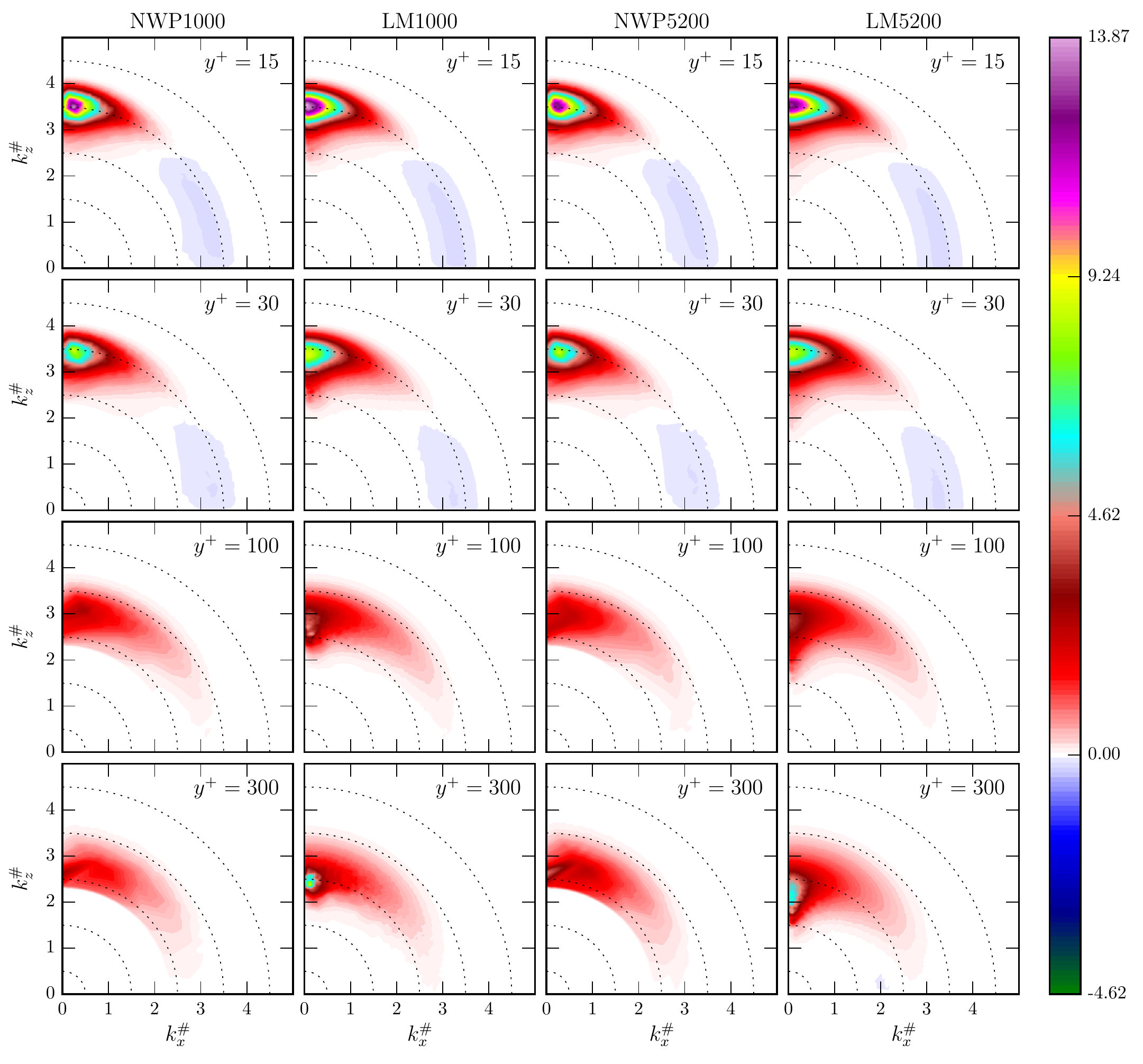}
\caption{Two-dimensional, premultiplied spectra 
$y^+\, (E^{\Pro}_{11})^+$ in log-polar
coordinates, as defined by equation \eqref{eq:log_polar_defn}. 
$\lambda^+=10$ on the outermost dotted circle and increases
by a factor of 10 for each dotted circle moving inward,
where $\lambda = 2\pi/k$ is the wavelength.}
\label{fig:2d_P_uu_wufl}
\end{center}
\end{figure}

The production of turbulent kinetic energy in a wall bounded flow is primarily 
due to the large mean velocity gradient in the wall normal direction $\partial U/\partial y$. 
In a flow that is homogeneous in the stream/spanwise directions with $V=W=0$, 
the only $\chevron{u'_{\alpha}u'_{\alpha}}$ term with a nonzero production is 
$\chevron{u'u'}$; it is given by 
\begin{equation}\label{eq:uu_production}
\Pro_{11} = -2\,\frac{\partial U}{\partial y}\, \chevron{u'v'}.
\end{equation}
The two dimensional spectra of $\Pro_{11}$ is accordingly defined as
\begin{equation}\label{eq:production_spectra}
E_{11}^{\Pro}(k_x,y,k_z) := -2\, \frac{\partial U}{\partial y}(y)\, E_{12}(k_x,y,k_z).
\end{equation}
The spectral analysis of channel flow data in \citet{Lee:2019}
 demonstrated that in contrast
to the near wall energy spectra $E_{11}$, the near wall production spectra 
$E^{\Pro}_{11}$ contains only a high wavenumber peak (see columns two and 
four in figure \ref{fig:2d_P_uu_wufl}).
It follows that the large scales in the 
near wall region, and hence the energy that they contain, 
 are due to energy transport (either in $y$ or in scale), rather than 
local production. This observation suggests that the  
NWP model should be able to capture the near wall energy 
production, even though the VLSMs are not present. The 
production spectra shown in figure \ref{fig:2d_P_uu_wufl} 
show that this is indeed true. 
At both $y^+=15$ and $y^+=30$, the NWP1000 and NWP5200 spectra are  
qualitatively similar to that of DNS, including
the regions of negative production occurring over a range
of scales around $\lambda^+=100$. 
Farther away from the wall, the large scale structures
increasingly influence the energy production, 
and their influence increases with $Re_{\tau}$. 
At $y^+=300$, the large scale influences dominate the 
DNS production spectra, and the model is not able 
to reproduce such low wavenumber features. 

\begin{figure}
\begin{center}
\includegraphics[width=0.5\textwidth]{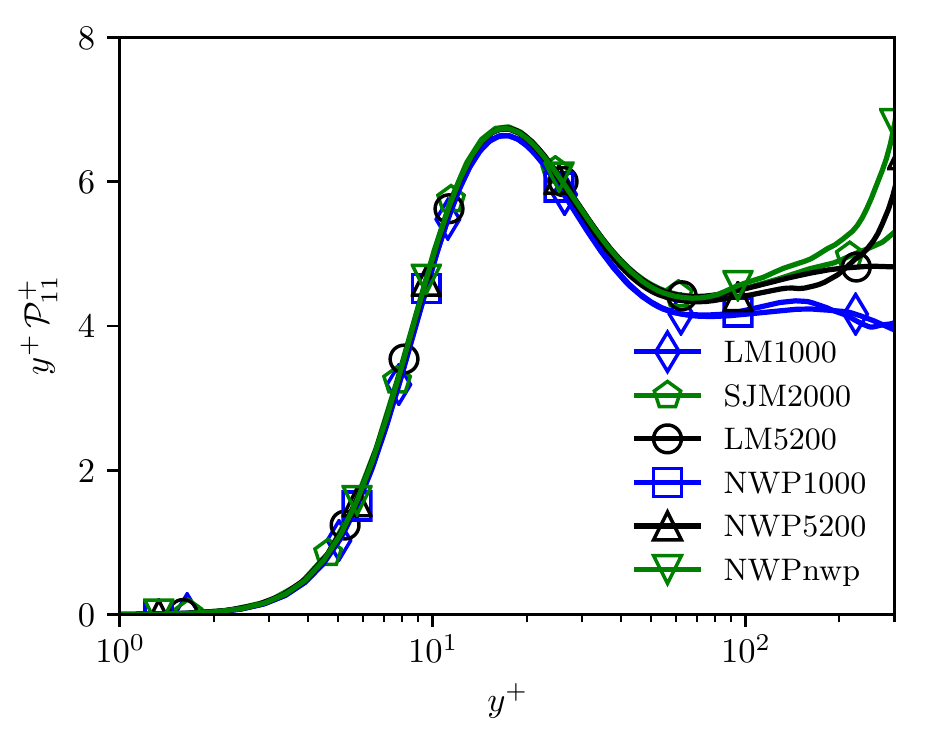}
\caption{Profiles of the premultiplied $\chevron{u'u'}$ production $\Pro^+_{11}$ 
 versus $\log(y^+)$ in the region $y^+\in[0,300]$.}
\label{fig:1d_production_wufl}
\end{center}
\end{figure}

The one-dimensional, premultiplied production profiles
are shown in figure \ref{fig:1d_production_wufl}, and they 
are consistent with the aforementioned observations regarding
the two-dimensional spectra. 
The DNS profiles are approximately 
$Re_{\tau}$-independent for $y^+\lesssim 70$, the corresponding
model profiles show strong agreement for $y^+\lesssim 100$, and 
they begin to show modest discrepancies for $y^+\gtrsim 200$. 


\begin{figure}
\begin{center}
\includegraphics[width=1.0\textwidth]{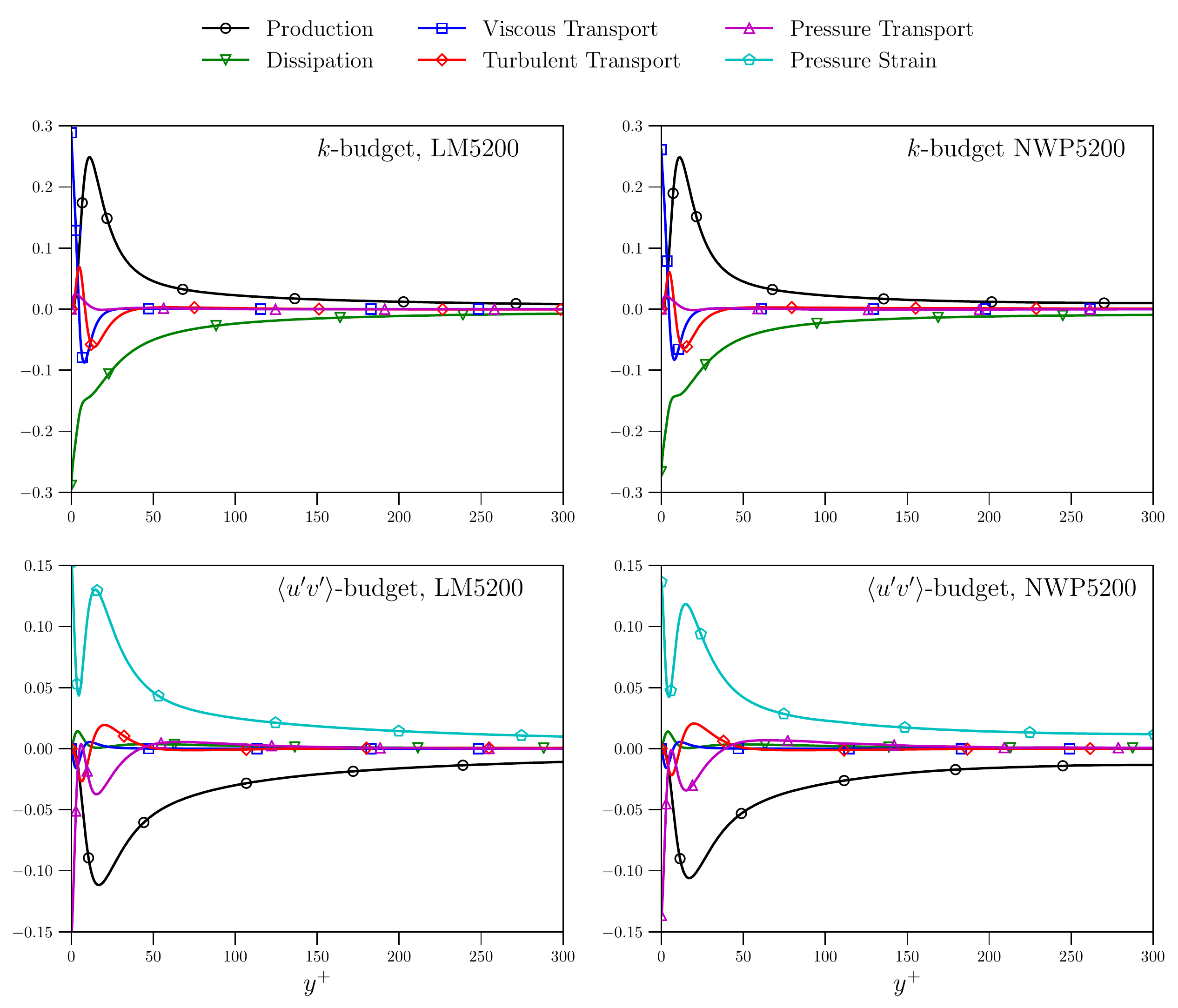}
\caption{
Budget equations of the turbulent kinetic energy
$k= \chevron{u_i'u_i'}/2$ (top) and shear stress $\chevron{u'v'}$ for
the largest Reynolds number cases LM5200 (left) and NWP5200
(right). The terms are defined in the
appendix \eqref{eq:Re_stress_transport_eq}
}
\label{fig:k_and_uv_budget}
\end{center}
\end{figure}

After being produced by the mean velocity gradient, turbulent kinetic energy is 
redistributed across scales and velocity components, transported
both towards and away from the wall, and ultimately dissipated
by viscosity. The relative strength, or importance, of these
processes as a function of wall-normal distance can be measured
by the terms in the Reynolds stress budget equation 
\citep{Pope:2000tp}.
 Exhaustive analyses of 
the behavior of these terms for wall bounded flows can be found in 
\citet{Hoyas:2008jl,Richter:2015ipa,Mizuno:2015wz,Mizuno:2016kx,Aulery:2016cg,Lee:2019},
and other references therein. A general conclusion to be drawn from these
works is that, similar to the production and velocity variances, the small
scale contributions to the terms in the budget equation are 
universal in the near wall region, and differences in the profiles as a 
function of $Re_{\tau}$ can be attributed to modulations by large
scale motions. 

As a consequence, the terms in the Reynolds stress budget equations
evaluated in the NWP model are consistent with those in near-wall
turbulence. For example, the terms in the turbulent kinetic
energy and Reynolds
shear stress budget equations are shown in
figure~\ref{fig:k_and_uv_budget} for both NWP5200 and LM5200.
The precise definitions of the terms in figure \ref{fig:k_and_uv_budget}
are standard, but 
for clarity they are specified in the appendix \eqref{eq:Re_stress_transport_eq}.

\subsection{Near-wall Turbulence Structure}
\begin{figure}
\begin{center}
\includegraphics[width=1\textwidth]{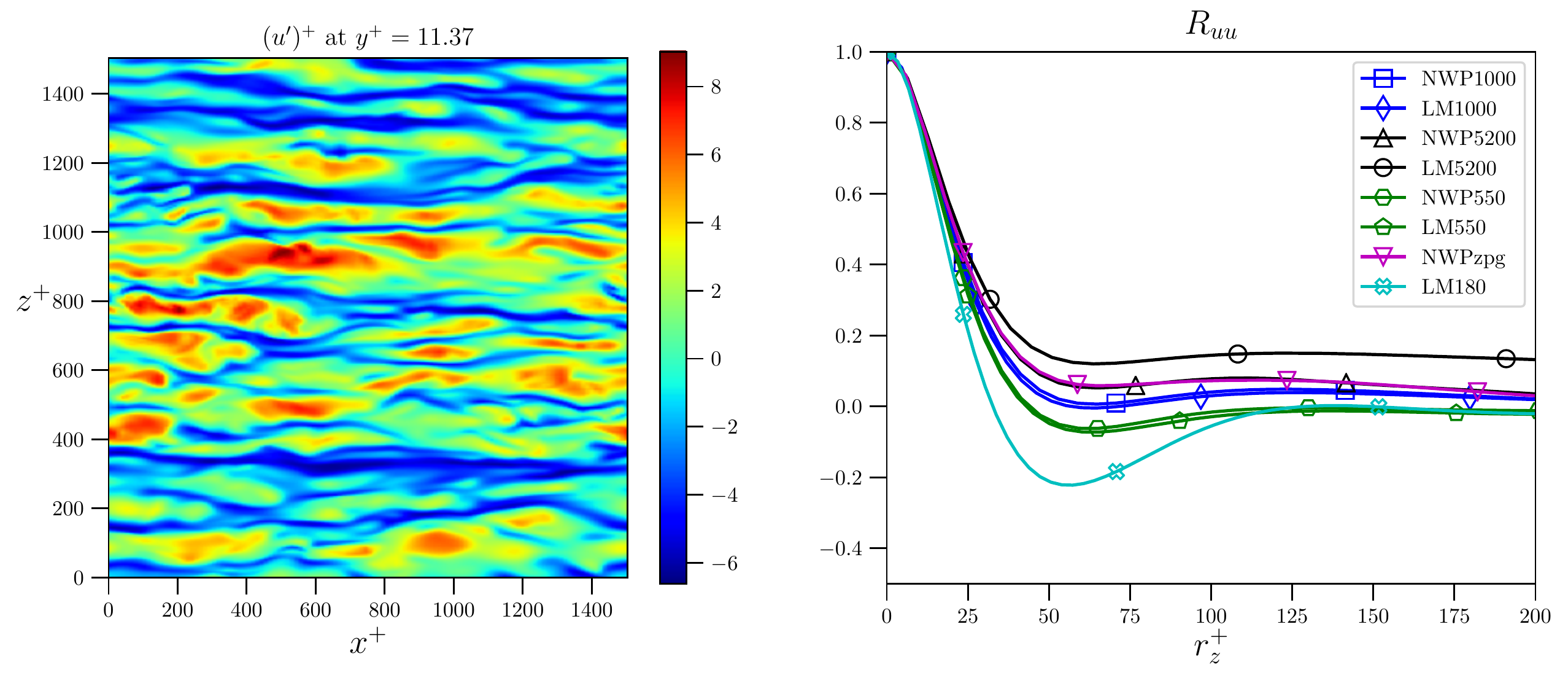}
\caption{
Left: snapshot of the streamwise velocity fluctuation $u'$ in the $x-z$ plane; 
right: two-point correlation function $R_{uu}(r_z)$ for all three model 
cases, as well as an additional model case NWP550, and 
channel flow DNS ranging from $Re_{\tau} \approx 
180-5200$. Each $R_{uu}$ is measured at $y^+\approx 11.5$. 
}
\label{fig:uprime_Ruu}
\end{center}
\end{figure}

In addition to producing consistent near-wall statistics, the NWP
model flow produces flow structures that are consistent with
expectations for near-wall turbulence. For example, the well-known
high- and low-speed streaks are apparent in visualizations of the
streamwise velocity fluctuations near the wall (figure~\ref{fig:uprime_Ruu}),
and those streaks have the expected spanwise spacing, as evidenced by
the two-dimensional spectra in figure~\ref{fig:2d_uu_wufl} at $y^+=15$, and
the spanwise two-point correlations of streamwise velocity
fluctuations in figure~\ref{fig:uprime_Ruu}. In the NWP flows, increasing the
magnitude of the imposed favorable pressure gradient appears to
increase the coherence of the streaks, as indicated by the depth of
the mild local minimum at $r_z^+\approx 50$, though the imposed
pressure gradient in the NWP5200 case is weak enough to have no effect
relative to NWPzpg. The correlations from NWP1000 and NWP550 are also
in excellent agreement with those of their corresponding channel flow
DNS, while the correlations from NWP5200 and LM5200 are quite
different. This is almost certainly due to the large scale modulations
of the near wall turbulence that occurs in the high Reynolds number
channel flow. This is apparent from the two-dimensional spectrum for
LM5200 at $y^+=15$ (see figure~\ref{fig:2d_uu_wufl}), which cannot be
represented in the NWP model, and which are largely absent in the
lower Reynolds number channel flows (LM1000 at $y^+=15$ in
figure~\ref{fig:2d_uu_wufl}).

\begin{figure}
\begin{center}
\includegraphics[width=1\textwidth]{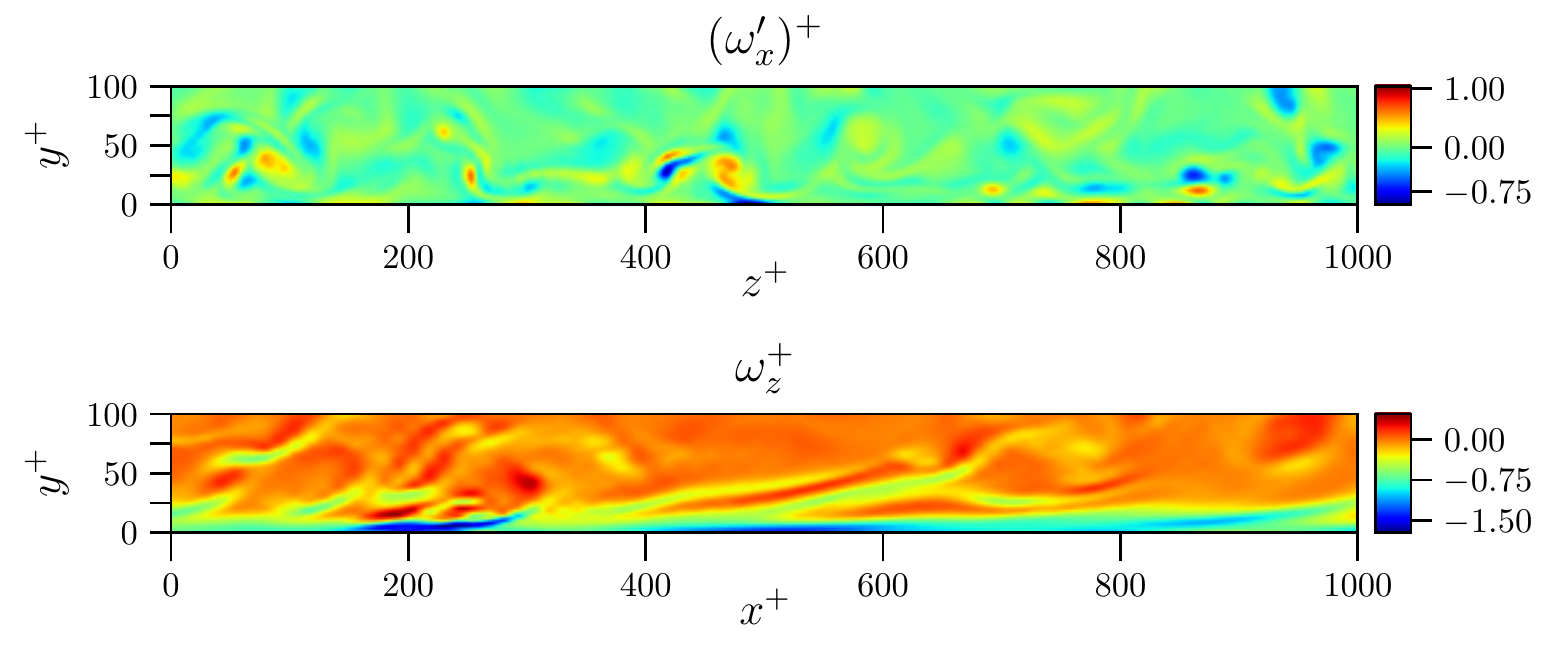}
\caption{Contour plots of the streamwise vorticity 
fluctuation $\omega_x'$ in the $y-z$ plane and spanwise vorticity 
$\omega_z$ in the $x-y$ plane.  }
\label{fig:omegax_omegaz}
\end{center}
\end{figure}

\begin{figure}
\begin{center}
\includegraphics[width=1.0\textwidth]{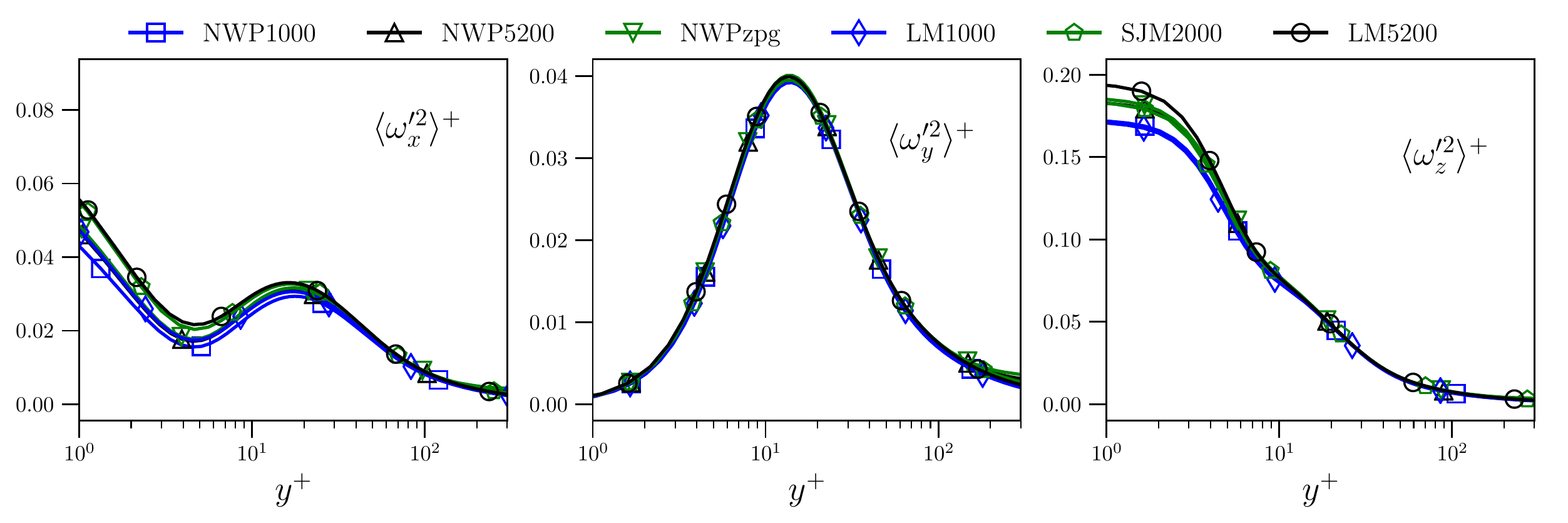}
\caption{
Vorticity variances $\chevron{\omega'_{\alpha}\omega'_{\alpha}}^+$ versus $\log(y^+)$
in the region $y^+\in[0,300]$. 
}
\label{fig:vort_variances}
\end{center}
\end{figure}

In the autonomous near wall dynamics, the near wall streaks are formed
by near wall streamwise vortices \citep{Jimenez:1999wf}, and such streamwise
vortices are indeed present in the NWP flow
(figure~\ref{fig:omegax_omegaz}). The presence of near wall
streamwise vortices is also imprinted in the streamwise vorticity
variance profiles (figure~\ref{fig:vort_variances}) in the local
maximum that occurs at about $y^+=15$. Streamwise, wall-normal and
spanwise vorticity variances in the NWP model flows are also all
largely consistent with those from the channel flow and boundary layer
DNS. Other features of the autonomous near wall dynamics include the
predominantly streamwise vortices that lift up from the wall forming
inclined vortices, as well as sharp gradients in the streamwise velocity that
are manifested as inclined spanwise vorticity structures. These too
are apparant in visualizations of spanwise vorticity from the NWP
flows (figure~\ref{fig:omegax_omegaz}).

\begin{figure}
\begin{center}
\includegraphics[width=1.0\textwidth]{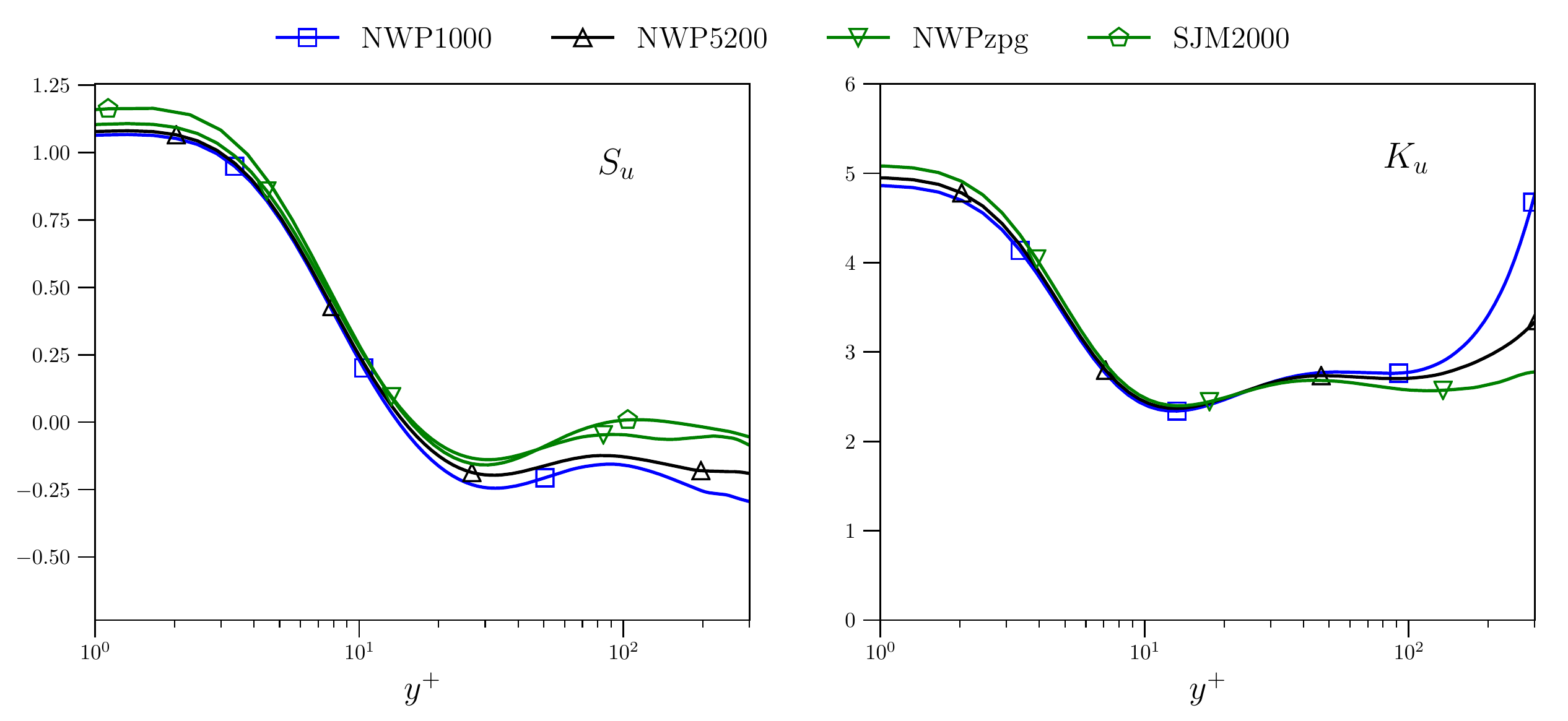}
\caption{
Skewness $S_u$ and kurtosis $K_u$ of the streamwise velocity fluctuations in the
NWP cases, and skewness in SJM2000.
}
\label{fig:skew_kurt}
\end{center}
\end{figure}

Given the consistency of the NWP flow structures with those of
near-wall turbulence, one would expect that higher order statistical
quantities would also be consistent. As an example, the skewness and
kurtosis of the streamwise velocity fluctuations are shown in
figure~\ref{fig:skew_kurt}. While these quantities are not as widely
reported, and are in particular not currently available for LM1000 and
LM5200, their profiles in the NWP cases are none-the-less consistent
with data in the literature \citep{monty2009,chin2015,Samie:2018}. In
particular, skewness has a local minimum at $y^+\approx 20$, crosses
zero at $y^+\approx 10$ and attains a maximum at the wall of about
1.1. Also, kurtosis has a local minimum at $y^+\approx 10$ and reaches
a maximum at the wall of about 5.



\section{Conclusions}
The near-wall patch (NWP) model of near-wall turbulence described here
was formulated in part to address the hypothesis that the autonomous
dynamics of the near-wall turbulence modulated by the large-scale
outer-layer turbulence is responsible for the observed characteristics
of near-wall turbulence. The NWP flow was also formulated to serve as
as a computationally accessible quantitative model of near-wall
turbulence which will be useful in a number of contexts.

Regarding the former objective, the study described here builds on
previous work by \citet{Jimenez:1999wf} and \citet{Moin:1991}, which
use simulations with restricted periodic domains, and in the case
of \citet{Jimenez:1999wf}, manipulation of the turbulence outside the
near wall layer. These simulations were used to characterize in detail
the fluid dynamic processes responsible for near-wall turbulence, and
establish that these processes are autonomous, not requiring
interaction with the outer turbulence. Here, however, the objectives
were different, and so the domain sizes, while still restricted, are
larger in the horizontal and vertical directions (1500 and 600 wall
units, respectively), motivated by the spectral analysis
of \citet{Lee:2019}).

The statistical profiles from the NWP flows are in close agreement
with those from DNS of turbulent channels and a boundary layer in
large spatial domains. For some quantities for which very large scales
make a significant contribution, such as the streamwsie velocity
variance, this agreement is attained only after the large-scale
fluctuations that are too large to be represented in the NWP have been
filtered out. This indicates that there is a universal near-wall
small-scale dynamics that produces the familiar statistics of
near-wall turbulence, which had been suggested previously based on
analysis of both experimental and DNS data from wall bounded turbulent
flows \citep{marusic2010,mathis_hutchins_marusic_2011,Lee:2019}. By
actually simulating the autonomous near-wall dynamics over the range
of scales at which they occur, as determined from spectral
analysis \citep{Lee:2019}, we confirm that the ``universal signal''
described by \citet{marusic2010,mathis_hutchins_marusic_2011} arises
from universal dynamics, where universal here means independent of
Reynolds number or external flow configuration.

As a quantitative model of near-wall turbulence, the NWP flow defines
a one-dimensional family of near-wall turbulent flows, parameterized
by the imposed streamwise pressure gradient $\partial P^+/\partial
x^+$. In this context, the NWP model can, for example, be used as a
source of data to inform a wall model for wall-modeled large eddy
simulation (LES). For such an application, one would invoke the scale
separation assumption discussed in \S\ref{sec:motivation} and use NWP flows
matched to the local pressure gradient and momentum flux associated
with the large-scale outer-layer flow simulated by the LES. 
Other
applications of the NWP model are as a vehicle for numerical
experiments on near wall turbulence as in \citet{Jimenez:1999wf} and
to investigate the interactions of the small-scale, near wall
turbulence with such complications as surface roughness, heat
transfer, chemical reactions, and turbophoresis. 
These applications of the
NWP were out of scope for the current study, but
the NWP model allows such near wall phenomena to be studied computationally at 
a much lower cost than a full DNS of a real turbulent flow. For example, 
the NWP computational grid is a factor of 24080 smaller than the DNS grid
used for the LM5200 channel case.

Finally, the NWP formulation described here can be considered a lowest order
asymptotic description of near-wall turbulence, in which the imposed
pressure gradient, the momentum flux from the outer flow and the mean
wall shear stress are considered uniform in space and time on the
scale of the patch. A higher order approximation could allow one or
more of these quantities to vary slowly in the streamwise direction or
time, which can be treated asymptotically as in \citet{Spalart:1988}
or \citet{topalian2017} to model the resulting spatial or temporal evolution of
the wall turbulence. This would broaden the applicability of NWP
models, which, for example, would allow them to be used with stronger
pressure gradients, especially strong adverse pressure gradient boundary
layers in which wall shear stress generally evolves relatively rapidly.

\section{Acknowledgments}
The work presented here was supported by the National Science Foundation Award
no. (DMS-1620396), as well as by the Oden Institute for Computational 
Science and Engineering. 
The research utilized the computing resources of the Texas Advanced Computing Center
(TACC) at the University of Texas at Austin. 

We wish to thank Myoungkyu Lee for assistance with the PoongBack code, supplying 
post-processing scripts for visualizing data, and insightful discussions. We also thank 
Todd Oliver, Prakash Mohan, and Gopal Yalla for insightful discussions and 
useful suggestions for this manuscript. 

The authors report no conflicts of interest.

\appendix
\section{Reynolds stress budget equation}
\label{sec:appendix_A}
The Reynolds stress budget equations govern the evolution of the Reynolds stress
tensor. The terms of the equation 
are given by: 
\begin{align}\label{eq:Re_stress_transport_eq}
\frac{D \chevron{u_i'u_j'}}{Dt} = &- \overbrace{\left(\chevron{u_i'u_k'}\frac{\partial U_j}{\partial x_k}
+\chevron{u_j'u_k'}\frac{\partial U_i}{\partial x_k}\right)}^{\mathcal{P}_{ij}} \\
&- \overbrace{\frac{\partial \chevron{u_i'u_j'u_k'}}{\partial x_k}}^{T_{ij}} 
+ \overbrace{\nu \frac{\partial^2\chevron{u_i'u_j'}}{\partial x_k\partial x_k}}^{D_{ij}}
\nonumber \\
&+ \overbrace{\chevron{p'\left(\frac{\partial u_i'}{\partial x_j}+\frac{\partial u_j'}{\partial x_i}\right)}}^
{\Pi_{ij}}
- \overbrace{\left(\frac{\partial \chevron{p'u_i'}}{\partial x_j} +
 \frac{\partial \chevron{p'u_j'}}{\partial x_i}\right)}^{\Upsilon_{ij}}
\nonumber \\
&- \overbrace{2\nu \chevron{\frac{\partial u_i'}{\partial x_k}\frac{\partial u_j'}{\partial x_k}}}^
{\epsilon_{ij}} \nonumber . 
\end{align}
Here $\mathcal{P}_{ij}$ denotes turbulence production, 
$T_{ij}$ turbulent transport, $D_{ij}$ viscous transport, 
$\Pi_{ij}$ pressure strain, $\Upsilon_{ij}$ pressure transport, 
and $\epsilon_{ij}$ dissipation.


\bibliographystyle{./jfm}

\bibliography{./new_reference}

\end{document}